\documentclass[12pt]{article}
\usepackage[a4paper]{geometry}
\usepackage[T1]{fontenc}
\usepackage[utf8x]{inputenc} 
\usepackage[affil-it]{authblk}
\usepackage{lmodern}
\usepackage[english]{babel}
\usepackage{amsfonts}
\usepackage{amssymb}
\usepackage{stmaryrd}
\usepackage{textcomp}
\usepackage{bbold}
\usepackage{amsmath}
\usepackage{authblk}
\usepackage{subfig}
\usepackage{caption}
\usepackage[toc,page]{appendix}
\usepackage{bbm} 
\usepackage{hyperref}
\usepackage{mathrsfs}
\usepackage[all]{xy} 
\usepackage{xcolor}
\usepackage{tikz} 
\usepackage{graphicx}
\usepackage{wrapfig}
\usepackage{lscape}
\usepackage{rotating}
\usepackage{epstopdf}
\usepackage{babel}
\usepackage{amsthm}
\usepackage{verbatim}
\usepackage{float}
\usepackage{setspace}
\usepackage{listings}
\PrerenderUnicode{é}
\usepackage{amsmath}

\date{September 5th 2017}
\title{An Empirical Approach to Financial Crisis Indicators Based on Random Matrices}
\author[1]{Raphael Douady}
\author[2]{Antoine Kornprobst  \thanks{Corresponding author: \texttt{antoinekor9042@gmail.com}}}
\affil[1]{Stony Brook University, Université Paris 1 Panthéon-Sorbonne}
\affil[2]{Université Paris 1 Panthéon Sorbonne\\ Labex ReFi}

\begin{document}
\maketitle
\begin{abstract}
The aim of this work is to build financial crisis indicators based on spectral properties of the dynamics of market data. After choosing an optimal size for a rolling window, the historical market data in this window is seen every trading day as a random matrix from which a covariance and a correlation matrix are obtained. The financial crisis indicators that we have built deal with the spectral properties of these covariance and correlation matrices and they are of two kinds. The first one is based on the Hellinger distance, computed between the distribution of the eigenvalues of the empirical covariance matrix and the distribution of the eigenvalues of a reference covariance matrix representing either a calm or agitated market. The idea behind this first type of indicators is that when the empirical distribution of the spectrum of the covariance matrix is deviating from the reference in the sense of Hellinger, then a crisis may be forthcoming. The second type of indicators is based on the study of the spectral radius and the trace of the covariance and correlation matrices as a mean to directly study the volatility and correlations inside the market. The idea behind the second type of indicators is the fact that large eigenvalues are a sign of dynamic instability. The predictive power of the financial crisis indicators in this framework is then demonstrated, in particular by using them as decision-making tools in a protective-put strategy.
\end{abstract}


\section{Introduction}
The objective of this paper is to build financial crisis indicators capable of producing a useful forecast of future market events. The goal that we set for this study is \textit{not} to predict the actual occurrence of financial crises. What we aim to achieve is rather to be able to evaluate at a given date whether the probability of a financial crisis happening at the given time horizon is getting higher, because the market conditions are ripe for a random adverse event from inside or even outside the market, to trigger a destructive chain reaction. Examples of random events capable of triggering a financial crisis are many. It may take the form of the sudden failure of a critical company, the publishing of new macro-economic data, a sovereign state defaulting on its debt, a major political event or even a terrorist attack. To use an analogy, we do not pretend to be able to predict the exact moment when a random spark will ignite the gas in the room, but we can measure whether the gas concentration in the room is just right for a random spark to cause a disaster. Since random adverse events happen all the time, measuring whether the conditions are just right in the market for one such event to trigger a crisis should be statistically equivalent to forecasting the actual occurrence of financial crises.\\

We build nine original financial crisis indicators which are divided into two kinds: those that study the distribution of the whole spectrum of the covariance matrix and compare it to a reference distribution and those that compute a specific spectral property (namely the trace and the spectral radius) of the covariance, correlation and weighted correlation matrix. Both kinds of indicators rely on the study of the underlying correlation and volatility signals inside the market. This is a novel approach because, while many different kinds of financial crisis indicators do exist in the literature, we are not aware of any that use reference distributions to compare the empirical spectrum of the covariance matrix to, nor any that use a modified version of the correlation matrix where the assets have been weighted with respect to the market capitalization of the corresponding companies or the daily traded volume. This approach enables us to maximize the amount of information coming from the market that is used by the financial crisis indicators, with the goal of boosting their predictive power. We work  with seven datasets, each one designed with its own unique composition characteristics. This provides us with original results about many different financial markets from North America to emerging countries.\\

There is a large literature on financial crisis forecasting, especially works by Sornette (2009), Sornette and Johansen (2010) , Jiang et al. (2010)  and Maltritz (2010) , which aim at producing a comprehensive model comprising the genesis, dynamics and eventual prediction of financial crises, especially using the powerful tools of time-series analysis. Network theory has also been successfully applied to financial crisis forecasting and the building of financial crisis indicators as in Celik and Karatepe (2007) or Niemira and Saaty (2004). A machine learning approach, based on K-means clustering, to forecasting financial turmoil, and especially sovereign debt crises, has been developed in Fuertes and Kalotychou (2007) who also demonstrated that combining multiple forecasting methods improves the quality of the predictions, as Clemen (1989) had underlined in a review and annotated bibliography about combining forecasts. Cross sectional time series analysis in a panel data framework was studied in Van den Berg et al. (2008) to predict financial crises while Bussiere and Fratzscher (2006) chose to develop early warning systems of financial crises based on a multinomial logit model. Demyanyk and Hasan (2010) summarized the results provided by several prediction methods of financial crises, and especially bank failures, based on economic analysis, operations research and decision theory, while Drehmann and Juselius (2014) proposed detailed evaluation criteria of the performance of early warning indicators of banking crises. Financial crisis forecasts can also be based on the quantitative study of any kind of qualitative macro-economic data like the FOMC \footnote{Federal Open Market Committee, which is the branch of the Federal Reserve Board that determines the direction of monetary policy} minutes, or any other qualitative forecasts. That approach was developed by Stekler and Symington (2016) as well as Ericsson (2016). Its main limitation resides in the quality of the qualitative forecasts and the FOMC for example did not predict the 2007-2008 financial crisis in advance nor did it identify it quickly as a major systemic event. From another point of view, Guégan (2008) used chaos theory and data filtering techniques to make market forecasts. The approach that we adopt is more modest in the sense that we do not pretend to explain the precise macro-economic mechanism that creates the many different kinds of financial crises and to predict the precise date of the next crisis. The ambition of this work is merely to detect a heightened risk of a crisis happening, not to predict its actual occurrence. The approach we adopt is closer to the work of Sandoval Junior and De Paula Franca (2012)  who proved in their paper, using random matrix theory techniques, that high volatility in financial markets is intimately linked to strong correlations between those financial markets.\\

Nonetheless, Sandoval Junior and De Paula Franca only used the Marchenko Pastur distribution in their work, while we intend to build and use additional distributions in the framework of random matrix theory. We also address internal correlations within the financial markets and not just the correlations between market indices. Those new distributions are numerically computed as closed form formulas for them do not exist to our knowledge. They are introduced in order to escape the restrictive framework of Marchenko-Pastur's theorem, which assumes uncorrelated Gaussian components. Indeed, the empirical covariance matrix of assets inside a market in turmoil is dominated by strong correlation and a non-Gaussian distribution of the log-returns. Of course, the objectives of this study are also very different, since we attempt to build empirical financial crisis indicators, which are almost ready for use by practitioners, while Sandoval Junior and De Paula Franca were concerned with proving a result about volatility and correlation reinforcing their effects during a financial crisis.\\

The approach and methods used in this study are also close to the work of Bouchaud, Potters and  Laloux (2005 and 2009). Indeed, in their 2005 physics paper and 2009 review, they apply random matrix theory and principal component analysis to the financial context in order to anticipate market events and produce optimal portfolio allocations as well as risk estimations. Their idea to use, like  Sandoval Junior and De Paula Franca, the Marchenko Pastur distribution as a reference distribution to which they compare empirical spectra is similar to the framework that we have developed but they use an exponentially weighted moving averages in place of the rolling matrix that we work with. The work of Singh and Xu (2013) and of Snarska (2007) about the dynamics of the covariance matrix in a random matrix theory framework was also inspirational to us. Indeed, the approach we select uses as well rolling windows for dynamic correlation and covariance matrices. Exploiting the spectra of those matrices forms the very foundation of the framework of this study.\\

We can also see the financial crisis indicators that we build as \textit{market instability indicators}. Indeed, they are able to say at a given date whether the probability of occurrence of a financial crisis within a given time horizon has increased, while it is still possible that the probability of nothing happening remains very high. In particular, one possible limitation of our approach is the relatively high ratio of false positives. There is still usually a high probability that nothing will happen, even when the indicators return red flags. From a practitioner's point of view, the information that the probability of a crisis occurring in the near future has risen from, say, 0.1\% to 10\% has tremendous value, even though there is still a 90\% chance of nothing happening. For us, a financial crisis indicator is a tool that makes use of publicly available data to determine whether the market conditions, measured by taking into account both  correlation and volatility, are ripe in the market for a crisis event to happen.\\

The robust methods used in this paper are applied to an intuitive principle of financial economics: when correlations between asset returns increase and develop abnormal patterns, when volatility goes up, then something is not right inside the market and a financial crisis event might be around the corner. Any kind of market data can be used within the framework that we created. Depending on the order of magnitude and scope of the financial crises that we intend to forecast, we have the freedom to choose the geographical characteristics of the data. Indeed, we can use prices time series restricted to assets located in one given country, one region or the whole world. The nature of the data can also be freely defined depending on the nature of the crisis events that we plan on forecasting. Stock prices and equity index prices, as well as sector indices may be used to forecast stock market crashes. Foreign exchange (FX) spreads may be used to forecast primarily monetary crises, and the methods that we develop provide a complementary point of view to the work of Guégan and Ielpo (2011) who used time-series models to forecast monetary policy. However, we are not limited to any asset class. We may also use bond yields, commodity prices or credit default swaps (CDS) spreads. Finally, it is possible to choose the frequency of the data and adapt it to reflect the kind and scope of the financial crises that we aim at forecasting, the only limitation being data availability.\\

In this paper, we chose to mainly focus on global financial crises, most of which are well known to the general public and the he data \footnote{The data we use in this paper has been collected from Bloomberg and Yahoo Finance.} has been selected accordingly. The code has been written using Matlab and its various optional toolboxes . The reader is very much encouraged to apply the methods developed in this paper to their own datasets and to verify the reproducibility of their forecasting power to various kinds and scopes of financial crises using data from many different kinds of asset classes and of various frequencies. We look forward to feedback and comments.\\

We propose in this paper a new approach regarding early warning financial crisis indicators that we then illustrate using many different datasets of market data. We also demonstrate the ability of the methods that we develop to make out-of-sample predictions. From our point of view, it seems that no such work has been published before with the same objectives and methodology. Therefore, we cannot compare quantitatively in terms of accuracy and predictive power the results that we have obtained to other existing studies. The work of Bouchaud, Potters  and Laloux (2005 and 2009) uses a methodology that is similar to the one we chose, however we did not find detailed empirical results for their work, that would have been suitable for comparison in a robust way with the numerical results that we have obtained in this study.\\

Besides the present introduction and a general conclusion, the paper is divided into four parts. We first describe how we built, collected and processed the databases. Indeed, their quality and diversity constitutes a major part of the interest of the study we conducted. In a second part, we detail the methodology and then we build the financial crisis indicators. The third part is dedicated to the qualitative analysis of the results provided by the financial crisis indicators over the whole length of the datasets. Finally, in the fourth part, we demonstrate the predictive power of the approach we developed by selecting two of the best performing financial crisis indicators applied to the largest and most detailed dataset that we possess. After dividing the data between an in-sample and an out-of-sample period, we study in details the forecasting possibilities they provide, firstly by using fixed dates of known financial crises and then by quantitatively defining a financial crisis in terms of the crossing of a chosen maximum draw down threshold.

\section{The Data}

The data is constituted at each date of the log-returns with respect to the previous trading day, computed from open or close prices. The prices have been adjusted for dividends and splits beforehand. We have chosen daily data for this study because of easy access and faster numerical handling. Further studies may explore higher frequency data. The model that we develop requires the choice of a rolling window in order to compute the financial crisis indicators. In order to limit averaging effects and to have financial crisis indicators with enough responsiveness  to provide useful information to a practitioner, we chose the size of the rolling window to be 150 days in the past. This represents roughly six months of trading since we only take trading days into account. Using a relatively large rolling window means that the covariance matrix will be degenerate sometimes since there will be more observations than assets. This fact however is not going to be a problem because for the first type of indicators, the distance between the empirical distribution and the reference will be computed after truncating the empirical distribution around zero and making it stick to the reference in order to eliminate the contribution of the small eigenvalues. The motivations for this operation will be explained in the next section where the methodology that we use is explained in detail. For the second type of indicators, the presence of zeros, even quite a lot of them, in the spectrum will not change anything for the computation of the trace and spectral radius.\\

Seven datasets, each designed with its own unique properties and composition are considered in this study  :

\begin{itemize}
\item The first dataset (Dataset 1) is constituted of eleven stock indices representative of the Asian, European and American financial markets in order to obtain a picture of the global financial system. It is a pure equity dataset that is designed to capture contagion between major financial markets as a way to forecast financial crises. It contains the Nikkei225 (NKY, Japan), Hang Seng (HSI, Hong-Kong/China), Taiwan Stock Exchange Weighted Index (TWSE, Taiwan) for the Asian market, the DAX30 (DAX, Germany), FTSE100 (UKX, U.K), IBEX35 (IBX, Spain) for the European market, the SP500 (SPX, U.S.A), Russel3000 (RAY, U.S.A), NASDAQ (CCMP, U.S.A), Dow Jones Industrial Average (INDU, U.S.A), SP/TSX Composite Index (SPTSX, Canada) for the North American market. Dataset 1 spans from January 7th 1987 to February 5th 2015. In order to avoid contaminating the data with time differences which might create bias and spurious correlations, we matched at a same date $t$ the close price in Asia at $t$, the close price in Europe at $t$ and the open price in America (East Coast) at $t$. In the absence of intraday data, this appeared to be a reasonable choice. We considered only the trading days and because of the different holidays specific to each of the three markets considered (Asian, European and North American) and the requirement to keep only the trading days that were common to all the markets, the 252 trading days a year have been reduced to around 200 dates. Comparison with the other datasets (particularly  Dataset 3 and Dataset 4 below which do have around 250 entries a year since they are exclusively American and European, respectively) shows that this is not a major issue in practice.

\item The second dataset (Dataset 2) is constituted of sixteen assets. It contains all of the indices of Dataset 1, some commodity indices and some safe haven or cash equivalent securities toward which investors tend to turn in a time of crisis or impending crisis. It spans the same period as Dataset 1, from January 7th 1987 to February 5th 2015. The treatment of the data with regard to time differences between geographical regions and non-trading days is the same. On top of the content of Dataset 1, Dataset 2 includes: The London Gold Market Fixing Index (GOLDLNPM, U.K), the Philadelphia Stock Exchange Gold and Silver Index (XAU, U.S.A), Oppenheimer Limited-Term Government Fund Class A (OPGVX, U.S.A), Sugar Generic Future Contract (SB1, U.S.A), generic First Crude Oil WTI (CL1, U.S.A). The inclusion of precious metal indices, cash equivalent short-government monetary funds, representative agricultural as well as energy commodities (in the form of investable futures) is supposed to provide a longer fuse to the financial crisis indicators. As a matter of fact, when the market starts to overheat, investors may liquidate some of their equity positions but they will have to re-invest the cash somewhere and those cash equivalent securities are here to account for that. Since those safer, cash equivalent securities are in Dataset 2, we anticipate that the risk of a crisis happening will be detected sooner. Moreover, when the market is becoming unstable, one typically witnesses an increase in the correlations between commodity and energy securities (typically large oil companies stocks included in the indices). Since we included some investable commodity futures (like oil futures) in Dataset 2, we expect to capture that effect which is indicative of the appearance inside the financial market of the right conditions for a crisis to happen.

\item The third dataset (Dataset 3) contains twelve assets which are the SP500 index and its ten sector sub-indices (consumer discretionary, consumer staples, energy, financials, health care, industrials, information technology, materials, telecommunication services, utilities) plus a small capitalization index, the Russel 2000. This dataset should provide information about the inner workings of the SP500 and enable us to detect "American" crises (for example the Sub-Prime Crisis of 2007) sooner and with a higher precision than Dataset 1 or Dataset 2 which are global by design and include information about the contagion between the three largest financial markets (Asia, Europe, North America). However, since the North American market still leads the world of finance, it is to be expected that the actual crises anticipated by the use of either three of Dataset 1, Dataset 2 or Dataset 3 will be roughly the same. The inclusion in the mix of a small capitalization index is to try to take advantage of the fact that in the times leading up to a financial crisis, the small caps tend to overheat and form speculative bubbles while they become more and more correlated between themselves and stocks with larger market capitalization. Dataset 3 spans from September 13th 1989 to December 27th 2013.

\item The fourth dataset (Dataset 4) is the European counterpart of Dataset 3. It contains eleven assets : the Bloomberg European 500 Index (BE500) and its ten sector sub-indices, which are the same as for the SP500 (consumer discretionary, consumer staples, energy, financials, health care, industrials, information technology, materials, telecommunication services, utilities). As we did not find any European-wide equivalent to the Russel 2000, it does not include small caps however. It should enable us to better and sooner detect "European" crises like the E.U Sovereign Debt Crisis of 2010 while still containing enough information to detect all the other global financial crises. It spans from January 1st 1987 to December 27th 2013.

\item The fifth dataset (Dataset 5) is designed with the financial concept of \textit{flight to quality} in mind. Indeed, in the times preceding a financial crisis, the anxiety of market agents is building up and they tend to abandon equity positions in favor of safer investment grade treasury or corporate bonds. In that regard, the usual observed phenomenon is a positive correlation between equity and bonds in a bull market and a negative correlation between equity and bonds in a bear market. When the correlation between equity and bonds is becoming too high, this may be a sign that the bull market is about to burn itself out, that a bubble is about to burst, heralding the start of a financial crisis. Dataset 5 is built with the detection of that phenomenon in mind. It contains all of the data of Dataset 3 (SP500 index, its 10 sector indices and the Russel 2000 as a small capitalization index) plus a number of funds based on investment grade sovereign or corporate bonds. Much like Dataset 3, Dataset 5 is U.S market oriented and is therefore more suited to anticipate crises that originate from or directly affect the North American market. For the long government bonds we have : Wasatch-Hoisington U.S. Treasury Fund (WHOSX) and Thornburg Limited Term U.S. Government Fund Class A (LTUCX). For the corporate bonds we have selected Lord Abbett Bond Debenture Fund Class A (LBNDX) and  Vanguard Long-Term Investment-Grade Fund Investor Shares (VWESX) which have both enough AUM (Assets Under Management) to be systemically significant and have existed for a long enough time to be historically relevant. Dataset 5 contains therefore 16 assets and spans from September 13th 1989 to December 27th 2013.

\item The sixth dataset (Dataset 6) is constituted of 226 individual components of the SP500 index. Because of the evolution over time in the composition of the index, a balance had to be found between keeping a sufficient number of components and having enough historical data. It spans from January 17th 1990 to May 15th 2015. The Apple Inc (AAPL) stock was chosen as the reference with regard to filtering out non-trading days and whenever another element of data was unavailable (on rare seemingly random days it appears that some individual stocks were not traded or the data was unavailable) we carried over the last previously available value. We assumed that this manipulation would not compromise the overall quality of the data. Besides those considerations, a few stocks like for example Range Resources Corporation (RRC UN) and The Charles Schwab Corporation (SHCW UN) presented significant data gaps and were removed from the dataset. Since building a dataset with exactly 500 components of the SP500, taking in account the evolution in the composition of the index over time, proved an impossible task due to its complexity and the availability of the data (mergers, corporate spin-offs and private equity acquisitions would have had to be taken into considerations as well), we are aware of the fact that Dataset 6, especially when used in conjunction with financial crises indicators might suffer from survivorship bias. As a matter of fact, especially in the times leading up to a crisis, the failing companies drop below the capitalization threshold or are acquired by others while new healthier firms enter the index. We built Dataset 6 because, as we are going to see in the empirical section, working with whole indices and/or limited number of individual securities like in all the previous datasets we created (especially Dataset 3 which resembles a scaled down version of Dataset 6), tends to have an averaging effect on the correlations and renders the correlation signal too noisy and blurred to be useful as a crisis detection method. For reference, the Bloomberg tickers of all the stocks inside Dataset 6 are provided in appendix. Besides the daily close price, from which we derive the log-returns, that is contained in all the other datasets, Dataset 6 also includes daily volumes and market capitalization. Those extra variables will enable us later to add appropriate weights to the individual stocks in order to refine the computation of the indicators.

\item The seventh dataset (Dataset 7) is constituted of the SP500, the Russel 2000 index and ten indices from emerging markets : Buenos Aires Stock Exchange Merval Index (MERVAL, Argentina), Ibovespa Brasil Sao Paulo Stock Exchange Index (IBOV, Brasil), Mexican Stock Exchange Index (MEXBOL, Mexico), Moscow Exchange Composite Index (MICEX, Russia), Hong Kong Hang Seng Index (HSI, Hong Kong/China), Shanghai Stock Exchange Composite Index (SHCOMP, China), Jakarta Stock Exchange Composite Index (JCI, Indonesia), National Stock Exchange CNX Nifty Index (NIFTY, India), FTSE/JSE Africa All Share Index (JALSH, South Africa), Borsa Istanbul 100 Index (XU100, Turkey). It spans from September 22nd 1997 to May 12th 2015. The relatively shallow depth of this dataset, which in particular may render the study of the Asian crisis of the late 1990' more difficult, is due to gaps in data availability, especially for the Russian index that we decided to keep anyway due to its importance for the global commodity and energy markets. All those emerging indices were expressed in the local currency on Bloomberg and were therefore converted into U.S dollars. This conversion was very important when dealing with emerging economies where the exchange rate of the local currency against the U.S dollar can fluctuate wildly and violently especially in the times leading up to, and during a financial crisis. Unlike in advanced economies (we did not convert the European and Japanese indices into U.S dollars in Dataset 1 for example), the position of the currency of an emerging country against the U.S dollar is also highly correlated to the health of the local real economy. This idea was developed by Hawkins and Klau (2000) when they were working with the Bank of International Settlements: in emerging markets, financial crises are often preceded by overvalued exchange rates and  inadequate international monetary reserves.
\end{itemize}

Regarding the selection of the financial crisis events on a global scale (for use mainly with Dataset 1 and Dataset 2) or at least a regional scale (for use mainly with Dataset 3, Dataset 4, Dataset 5, Dataset 6 and Dataset 7) of the last 30 years, we compiled Table 1 below, which has no ambition of being exhaustive. Succinct  historical context will be discussed in the empirical results section when needed. While categorizing the various kinds of financial crises goes far beyond the scope of this paper, we strove to consider a wide selection in the kinds of crises. There are stock market crashes like Black Monday in 1987 and the NASDAQ Crash in 2000. There are financial crises that are rooted into a deep structural fragility of some parts of the real economy, like the real estate sector in the case of the Japanese Asset Price Bubble of the early 1990' and the Sub-prime Crisis that started in America during the summer of 2007 or the automobile industry in the case of the bankruptcy of General Motors in June 2009, four years after Delphi Corporation, which was General Motors' main supplier of automotive parts. There are financial crises for which the main trigger was a sovereign debt default like the Russian crisis in 1998, the Argentine crisis in 2001 or the Eurozone crisis, triggered by the Greek haircut in 2010. There are monetary crises as well, like Black Wednesday in 1992 when the British government was forced to withdraw the Pound Sterling from the European Exchange Rate Mechanism (ERM) or the Mexican crisis triggered by the devaluation of the peso against the U.S Dollar. Since none of the datasets include foreign exchange data, we do not expect that any of the indicators will perform well when it comes to anticipating monetary crises, however. There are banking crises as well such as the S\&L crisis in America that spanned from the mid-1980' to the mid-1990' and during which almost one third of all American savings and loans associations (financial institutions that are allowed to accept savings deposits and to make loans) failed, including hundreds of banks of all sizes and systemic significance. The dates chosen may sometimes seem a little arbitrary but choices had to be made, especially for crises that, unlike Black Monday that played out mostly within a few days of extreme market distress, took place over many months or even years of sustained drop like the NASDAQ in early 2000, which took nearly four months to lose almost two fifth of its March 10th peak. Most financial crises do not happen in one day and instead result from a long process of instability buildup inside the market, the kind of which the indicators that we have built are detecting. When a crisis is best described by a clear explosion, then the date of that event was chosen (Black Monday, the day Lehman Brothers failed, etc...). When a date for a crisis spanning months or years had to be chosen for this study, we considered either the date of the most marking event (the day the NASDAQ peaked, the day General Motors filled for Chapter 11 bankruptcy \footnote{Chapter 11 of Title 11 of the United States Code (also known as the United States Bankruptcy Code) which permits reorganization. In contrast, Chapter 7 provides a legal framework for liquidation.}, the day the Greek haircut was announced, etc...) or a date roughly situated in the middle of the crisis process like January 1st 1990 for the S\&L crisis.\\

\begin{center}
\begin{tabular}{ | l | l | l | }
\hline
	Date (Y/M/D) & Name \\ \hline
	1987-10-19  & Crisis 1 : Black Monday \\ \hline
	1990-01-01  & Crisis 2 : S\&L Crisis \\ \hline
	1990-08-01  & Crisis 3 : Japanese Asset Prices Bubble Burst \\ \hline
	1991-09-19  & Crisis 4 : Scandinavian Banking Crisis \\ \hline
	1992-09-16  & Crisis 5 : Black Wednesday \\ \hline
	1994-12-20  & Crisis 6 : Mexican Crisis \\ \hline
	1997-07-25  & Crisis 7 : Asian Crisis \\ \hline
	1998-08-17  & Crisis 8 : Russian Crisis \\ \hline
	2000-03-10  & Crisis 9 : NASDAQ Crash (dot-com Bubble) \\ \hline
	2001-02-19  & Crisis 10 : Turkish Crisis \\ \hline
	2001-09-11  & Crisis 11 : 911 Attacks \\ \hline
	2001-12-27  & Crisis 12 : Argentine Default \\ \hline
	2005-10-08  & Crisis 13 : Delphi (G.M) Bankruptcy \\ \hline
	2007-07-01  & Crisis 14 : Sub-prime Crisis \\ \hline
	2008-09-15  & Crisis 15 : Lehman Brothers Collapse \\ \hline
	2009-06-01  & Crisis 16 : General Motors Bankruptcy \\ \hline
	2010-04-23  & Crisis 17 : European Sovereign Crisis \\ \hline
	2011-08-05  & Crisis 18 : US Sovereign Credit Degradation \\ \hline
	2014-12-16  & Crisis 19 : Russian Financial Crisis \\ \hline
\end{tabular}
\end{center}
\begin{center}
(Table 1: Selection of Financial Crisis Events in the past 30 years)
\end{center}

\section{Methodology}
Using the seven datasets that we have built, the methodology is based on the use of the spectrum of the covariance matrix, the correlation matrix and a weighted version of the correlation matrix. At each date, for a sequence of rolling windows, we either compare the whole spectrum to three reference distributions detailed below (two of the reference distributions characterize a calm market and a third one represents a market in turmoil), which gives us the indicators of the first kind, or we merely compute the spectral radius and the trace, which gives us the indicators of the second kind. We now details this methodology using Matlab's formalism and vector indexing conventions. 

\subsection{Framework}
We decided to consider a rolling window $T$ of 150 days in the past at each date and for all datasets, irrespective of the number $N$ of assets they contain. This choice provides us with a good balance between the readability of the signals, favored by a longer rolling window because of the averaging effect, and the responsiveness of the indicators, favored by a shorter rolling window.\\

For each of the seven datasets, we build at each date $t$ a rolling window $ROL(t)$ of length $T$. Then, we compute the rolling covariance matrix $CV(t)$ and the rolling correlation matrix $CR(t)$ by using the following formulas written for every row (i.e asset) $j \in \llbracket1,N \rrbracket $: 
$$ROL^{*}(t)(j,:) = ROL(t)(j,:) - mean(ROL(t)(j,:)) ~~ (1)$$
$$CV(t) = \frac{1}{T}  . ROL^{*}(t) \times (ROL^{*}(t))^{'}~~ (2)$$ 
$$CR(t)(j,:) = \frac{ROL^{*}(t)(j,:)}{\sqrt{var(ROL(t)(j,:))}} ~~ (3)$$

While working with a covariance matrix instead of a correlation matrix, we of course have to rescale the eigenvalues of $CV(t)$. We perform this either by noticing that the standard deviation of financial log-returns is typically in the order of magnitude of a few percents ($a \in [0.01,0.03]$) and therefore multiplying the eigenvalues by $\frac{1}{a^{2}}$, or by computing the mean of the variances of all the complete time-series in advance and multiplying by the inverse of that value (for example, we find a rescaling factor of 3410 for Dataset 2). This is what we decided to do but it should not be considered as a violation of the measurability of the indicators with respect to the natural time filtration (i.e knowledge of the future). It is just a practical way of rescaling by choosing the most appropriate value and it could just as well have been obtained from historical data predating the sample.\\

With regard to the reference distributions we use for the first type of indicators, we have built three of them : 
\begin{itemize}

\item $\Theta1$: the theoretical Marchenko Pastur distribution. It is derived from Marchenko Pastur's theorem presented in the work of Marchenko and Pastur (1967). Let $X$ be a $N \times T$ random matrix of i.i.d normal $\mathscr{N}(0,\sigma^{2})$ coefficients (in this study, each row represents an asset and each column represents an observation at a date $t$), then when $N,T \longrightarrow \infty$ and the aspect ratio of the matrix, $N/T \longrightarrow \gamma < \infty$, then the distribution of the eigenvalues of the covariance matrix $Y = \frac{1}{c} (X  X^{'})$ is the Marchenko Pastur distribution with the density below. This formula (6) below is supposed to be valid for $0<\gamma < 1$, otherwise in the degenerate case, an atom at zero has to be added, but since we intend to truncate the computation of the Hellinger distance to exclude the very small eigenvalues, as we will explain in the next section, this is the formula we are going to use anyway for simplicity.
$$\lambda^{+} = \sigma^{2}(1 + \sqrt{\gamma})^{2}~~ (4)$$
$$\lambda^{-} = \sigma^{2}(1 - \sqrt{\gamma})^{2}~~ (5)$$
$$f (x) =  \frac{1}{2 \pi \sigma^{2}\gamma} \frac{\sqrt{(\lambda^{+} -x)(x - \lambda^{-})}}{x} 1_{[\lambda^{-},\lambda^{+}]}~~ (6)$$
The Marchenko Pastur distribution also provides thresholds $\lambda^{+}$ and $\lambda^{-}$ that we use even while working with other simulated reference distributions. Because of the stringent theoretical requirements of Marchenko Pastur's theorem, that will never be even remotely satisfied by real financial data, the Marchenko Pastur distribution has no vocation to be the best distribution fulfilling the role of a calm market reference, but it is still going to be useful in this study.

\item $\Theta2$: the distribution of the eigenvalues of the covariance matrix of a simulated random matrix made of Gaussian $\mathscr{N}(0,1)$ coefficients where the assets, materialized as the rows, present some correlation to one another. Let us consider a rolling matrix  $(Z_{i,j})_{(i,j)\in \llbracket0,N \rrbracket \times\llbracket0,T\rrbracket }$, containing  $T$ observations (columns) and $N$ assets (rows) and constituted of i.i.d Gaussian $\mathscr{N}(0,1)$ coefficients. We introduce correlation between the assets by adding the same Gaussian coefficient to each of the assets at a given time. $\forall j \in \llbracket 0, T\rrbracket $,  $Z^{j}_{0}$ is following a Gaussian $\mathscr{N}(0,1)$ law. With those notations, $\forall i \in \llbracket0,N \rrbracket$ and  $\forall j \in  \llbracket0,T \rrbracket$, each of the coefficients $X_{i,j}$ of the rolling random matrix from which we obtain the covariance matrix, is computed in the following manner: $$X_{i,j} = \rho Z^{j}_{0} + \sqrt{(1-\rho^{2})}Z_{i,j}~~ (7)$$ $$Z_{i,j} \sim \mathscr{N}(0,1), ~ Z_{0}^{j} \sim \mathscr{N}(0,1)~~ (8)$$ The coefficient $\rho$ is chosen as the mean of the long term correlation coefficients between all the assets of the whole sample contained in the chosen dataset. Like when we had to decide on a rescaling coefficient for the spectrum, this choice of $\rho$ should not be considered as knowledge from the future, as it could just as well have been obtained from historical data. As a matter of fact, we find something very close to 50 \% for all the datasets, which is what we had expected.  $\Theta2$ is going to be the second, more realistic, calm market reference distribution.

\item $\Theta3$: the distribution obtained using the same blueprint as $\Theta2$ but where all the Gaussian $\mathscr{N}(0,1)$ distribution have been replaced by Student (t=3) distributions. This will be the reference distribution characterizing a market in turmoil.
\end{itemize}
As an illustration, Figure 1 below contains the three reference distributions $\Theta$1, $\Theta$2 and $\Theta$3 for the number of assets contained inside Dataset 1 (eleven assets) and a rolling window of 150 days. We also included in appendix the three reference distributions  computed for all the other datasets and the same rolling window of 150 days.

\begin{center}
\includegraphics[width=115mm]{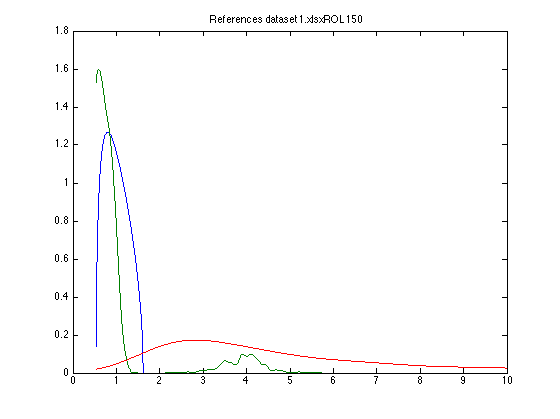}\\
Figure 1: Reference distributions for Dataset 1.  Blue: Marchenko Pastur ($\Theta$1), Green: $\Theta$2, Red: $\Theta$3
\end{center}

\subsection{Financial Crisis Indicators}
Using the tools described above, we build two kinds of financial crisis indicators: the indicators of the A-series and the indicators of the B-series.\\

The indicators of the A-series compare at each date the empirical distribution of the spectrum of the covariance matrix to the references that we introduced in the previous subsection. We chose to use the Hellinger distance in its discrete form as introduced by Hellinger (1909). We decided to use this metric on the space of distribution instead of the Kullback–Leibler divergence introduced in Kullback and Leibler (1951), which is also of very common use in probability theory, because we wanted a true metric, which the Kullback–Leibler divergence is not since it does not satisfy the triangle inequality.  The Kullback–Leibler divergence is also not symmetric with respect to the two distributions considered. Moreover, since we chose an empirical approach with a strong focus on the intuitive aspect of the study, we felt that the Hellinger distance, which is plainly the Euclidean distance of the square root of the components, was easier to see when drawing two distributions on the same graphic than the Kullback–Leibler divergence which is defined as the expectation of the logarithmic difference between the two distributions.\\

We recall that the Hellinger distance $ \mathbb{D}$ between two probability distributions with densities $P(x)$ and $Q(x)$, which are both known at a number of points $X_{i}$, ($i \in \llbracket 1, K \rrbracket $), is computed using the formula below :
$$ \mathbb{D}^{2} = \sum_{i=1}^{K}(\sqrt{P(X_{i})} - \sqrt{Q(X_{i})})^{2} ~~ (9)$$

Considering any of the three reference distributions, we compute at each date its Hellinger distance to the empirical distribution of the eigenvalues of the covariance matrix. Our assumption is that the further away in the sense of the Hellinger distance the empirical distribution drifts away from the calm market reference, and the closer in the sense of the Hellinger distance the empirical distribution comes to the market in turmoil reference, then the more likely it becomes that the market is about to experience a crisis. Indeed, such movements tend to indicate a build-up of correlation and volatility inside the market. There is however no way to study those two effects, volatility and correlation, separately in the Hellinger distance approach and the indicators of the A-series always lump those two instability factors together in their forecasts of financial crises.\\

Since all datasets except Dataset 6 only have a small number $N$ of assets and will therefore give us only a small number of eigenvalues at each date, we combine at each date $t$ the spectra obtained from the previous $20$ days in order to have $20.N$ eigenvalues, which is enough observations, to derive a distribution by using a normalized histogram. We then compute the Hellinger distance to the reference distribution on a sufficiently large support in order to capture all of the spectral distribution. In empirical studies such as Stanley and al. (2000), eigenvalues of the covariance matrix have been observed to grow as large as twenty five times the  critical value $\lambda^{+}$ of Marchenko Pastur's distribution (4). In consequence, we decided to consider 25 times the support  $[\lambda^{-}, \lambda^{+}]$ in order to account for all of the empirical spectrum.\\

The indicators of the first type are the following. There are three of them, corresponding to the three reference distribution that we have introduced before: 

\begin{itemize}
\item Indicator A1: It is the Hellinger distance between a modified version $\mathscr{E}_{1}^{*}(x)$, detailed below,  of the empirical distribution $\mathscr{E}(x)$ of the eigenvalues of the covariance matrix and the theoretical distribution of Marchenko Pastur $\Theta1$. Indeed, Indicator A1, as well as all the indicators based on the Hellinger distance, needed to be adapted to filter out the effects of a parasitic phenomenon consisting of an accumulation of small eigenvalues close to zero, which deforms the unmodified empirical distribution and distorts the computation of its Hellinger distance with respect to the reference distribution. We illustrate this in Figure 2 and Figure 3:

\begin{center}
\includegraphics[width=68mm]{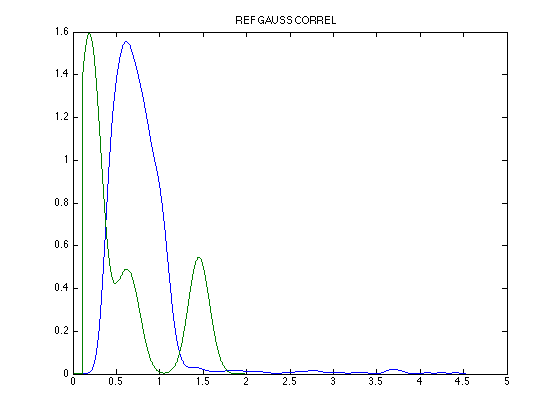} \\
Figure 2\\
Accumulation of Small Eigenvalues for $\Theta2$
\end{center}

\begin{center}
 \includegraphics[width=68mm]{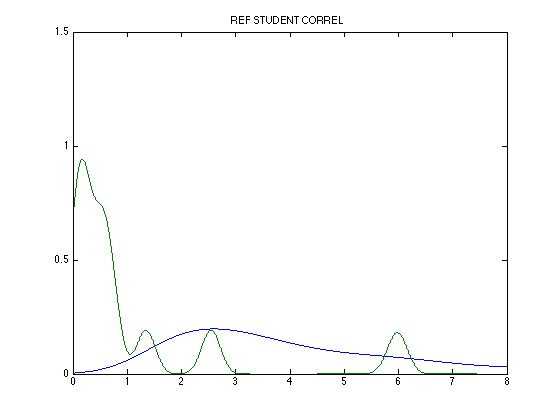}\\
 Figure 3\\
 Accumulation of Small Eigenvalues for $\Theta3$
\end{center}

In those two examples, which are very typical of the situations that we encounter in practice for a given date $t$ while using all the datasets, we see the reference distribution (in blue) and we see the empirical distribution of the eigenvalues of the covariance matrix (in green). The empirical distribution can differ from the reference distribution in two ways. It can overflow to the right toward the higher eigenvalues: that's the kind of behavior that we are looking for in order to detect a financial crisis. It can also unfortunately accumulate itself, sometimes most of the mass is even there, closer to zero toward the very small eigenvalues. Such a behavior of the distribution of the eigenvalues of the covariance matrix is more indicative of the prevalence of risk free combinations of assets which equates to a very calm and diversified market.

The solution was to define A1 and the A-series indicators in the following way : instead of computing the Hellinger distance between the unmodified empirical distribution $\mathscr{E}$ and the Marchenko Pastur reference $\Theta1$, we compute the Hellinger distance between $\Theta1$ and the distribution $\mathscr{E}_{1}^{*}$ defined in the following way : 

$$\mathscr{E}_{1}^{*}(x) = min(\mathscr{E}(x),\Theta1(x)), ~ x< \frac{\lambda^{+} }{10} ~~ (10)$$
$$ \mathscr{E}_{1}^{*}(x) =  \mathscr{E}(x), ~  x> \frac{\lambda^{+}}{10} ~~ (11) $$
Therefore : $$A1 =  \mathbb{D} \lbrace\mathscr{E}_{1}^{*} , \Theta1  \rbrace  ~~ (12) $$

For a given dataset, this indicator measures at each date by how much the assumptions of Marchenko Pastur's theorem (normal i.i.d coefficients of variance equal to 1) are violated. Since the finite size of the rolling covariance matrix does not change over time, it will not be responsible for any dynamical variations of the Hellinger distance although it does certainly account for part of the distance between the theoretical asymptotic distribution of Marchenko Pastur and the empirical distribution. This indicator lumps together the apparition of non-normality, correlations and volatility in the log-return time series, it cannot differentiate between all those effects but it is still very useful. As a matter of fact, the apparition of any of those phenomena, whose effects are not expected to compensate one another, can be interpreted as a warning that a crisis might be around the corner. Therefore our assumption is going to be that the further away the modified empirical distribution $\mathscr{E}^{*}(x)$ becomes from the reference Marchenko Pastur distribution in the sense of the Hellinger distance, the more likely a crisis is going to happen in the near future.\\

Like we said earlier, using the Marchenko Pastur distribution for a given aspect ratio as a reference distribution might not be optimal because it is an asymptotic result and we deal with finite size matrices and because even a perfectly calm financial market might be better modeled by random matrices of coefficients with some natural correlations.\\

\item Indicator A2: It is the Hellinger distance between the modified version $\mathscr{E}_{2}^{*}(x)$, detailed below,  of the empirical distribution of the eigenvalues of the covariance matrix $\mathscr{E}(x)$ and the simulated reference distribution $\Theta2$. Since correlated Gaussian coefficients are supposed to better model the market situation, we expect $\Theta2$ to provide a better calm market reference from which to measure a drift of the empirical distribution of the eigenvalues of the covariance matrix in the times leading up to a financial crisis. Like with Indicator A1, we decided to work on 25 times the support of the corresponding theoretical Marchenko Pastur distribution and the same issues of accumulation of the empirical distribution toward the small eigenvalues in time of market calm presented itself. There is no closed form formula for the reference $\Theta2$ and we cannot assume that its support is bounded like the support of Marchenko Pastur's distribution is bounded by  $\lambda^{-}$ and $\lambda^{+}$ so we decided to keep $\lambda^{*}=\frac{\lambda^{+}}{10}$ as a threshold, such that an abundance of very small eigenvalues would not make the Hellinger distance explode. Therefore, we compute the Hellinger distance between $\Theta2$ and  $\mathscr{E}^{*}$ in the following manner :
$$\mathscr{E}_{2}^{*}(x) = min(\mathscr{E}(x),\Theta2), ~ x< \lambda^{*} ~~ (13)$$
$$ \mathscr{E}_{2}^{*}(x) =  \mathscr{E}(x),  ~x \geq \lambda^{*}~~ (14)$$
Therefore : $$A2 =  \mathbb{D} \lbrace\mathscr{E}_{2}^{*} , \Theta2  \rbrace ~~ (15)$$

\item Indicator A3: It is the Hellinger distance between the modified version $\mathscr{E}_{3}^{*}(x)$, detailed below,  of the empirical distribution of the eigenvalues of the covariance matrix $\mathscr{E}(x)$ and the simulated reference distribution $\Theta3$. We included very fat tails (coefficients that follow a Student (t=3) distribution) as a way to model crisis conditions, therefore Indicator A3 is an inverted indicator. Indeed, it produces red flags when it is getting small, which means that the empirical distribution of the eigenvalues of the covariance matrix is getting very close to $\Theta3$, with is extremely heavy tailed and represents a spectrum entirely shifted toward the large eigenvalues, characterizing a market in deep turmoil. When the market goes from a calm state to a crisis state, the modeling that we make of the log-returns goes from a Gaussian to a Student (t=3) distribution. As a remark, we did not include skewness in the random coefficients from which we derive the reference distributions because financial log-returns do not typically present persistent skewness, especially over the time periods considered for the rolling window, as demonstrated in the work of Singleton and Wingender (1986). We retain for A3 the same method of computation as in the other ones of the A-series. We compute it over 25 times the support of the corresponding theoretical Marchenko Pastur distribution and the threshold $\lambda^{*}=\frac{\lambda^{+}}{10}$ is used to filter out the very small eigenvalues. Indicator A3 then computes at each date $t$ the Hellinger distance between $\Theta3$ and  $\mathscr{E}^{*}$ such that :
$$\mathscr{E}_{3}^{*}(x) = min(\mathscr{E}(x),\Theta3), ~ x<\lambda^{*} ~~ (16)$$
$$ \mathscr{E}_{3}^{*}(x) =  \mathscr{E}(x), ~ x \geq  \lambda^{*} ~~ (17)$$
Therefore : $$A3 =  \mathbb{D} \lbrace\mathscr{E}_{3}^{*} , \Theta3  \rbrace ~~ (18)$$
\end{itemize} 

We therefore have three indicators of the first type, called A1, A2 and A3. Each possesses its own characteristics and looks at specific market conditions that may be indicative of an impending financial crisis. We do expect the three indicators of the first kind to be coherent with one another, especially since they are of similar origin, but they also complement one another and the financial crisis forecasts that we make need to take all three into consideration to be effective.\\

We now shift to the indicators of the B-series. At each date $t$, the centered rolling matrix $ROL^{*}(t)$ in formula (1) contains two components: a volatility component and a correlation component. The indicators of the B-series are based on those two components. As we are going to see, both components are important, but the relative strength of their signal will greatly depend on the choice of the dataset we use. We build at each date $t$ the three indicators of the second type in the following way :
\begin{itemize}
\item Indicator B1: It is defined as the spectral radius of the covariance matrix $CV(t)$ in formula (2). It measures a mixed signal depending on both volatility and correlations in the market. A larger value for the spectral radius is indicative of dynamical instability and increased correlations in the system but it also takes the volatility effect into account since we are working with a covariance instead of a correlation matrix. This indicator takes the effects of both volatility and correlations into account and those two effects are not supposed to compensate each other, on the contrary they are expected to evolve in the same direction in the times leading up to a financial crisis as it was demonstrated by Sandoval Junior and De Paula Franca (2012).

\item Indicator B2: It is defined as the trace of the covariance matrix $CV(t)$. It measures the volatility signal alone. While it may seem at first that B2 lacks a very important aspect of what is happening inside the market, we will see while discussing experimental results that it is still a very good financial crisis indicator. It is also very easy and fast to compute. As a matter of fact it is not even needed to compute the whole spectrum of the covariance matrix to compute its trace.

\item Indicator B3: It is defined as the spectral radius of the correlation matrix $CR(t)$ in formula (3). It measures the correlation signal alone. The usefulness of Indicator B3 greatly depends on the choice of the dataset. We will discuss more about this in the section discussing the numerical results. Only when used on Dataset 6, which contains a large number of assets, which are individual stocks components of an index, does indicator B3 realize its full potential. Indeed, there is a lot averaging effect inside an index and when we use the value of the index itself as opposed to its individual components, the correlation signal is generally smothered. The potential of Indicator B3 is great however, because unlike with the study of volatility alone, the study of correlation may be the only way to give the indicators that we have built real predictive power. Since Dataset 6 also features daily volume and daily market capitalization data, we also build the following variations of indicator B3, to be used on Dataset 6 exclusively:
\begin{itemize}
\item Indicator B3A: the spectral radius of the matrix $CR_{1}(t)$. Its coefficients are those of $CR(t)$ which have been weighted at each date $t$ by the market capitalization ($cap(t)$) expressed in dollars, in the following way for a dataset containing $F$ assets. $\forall (i,j) \in [1,F]^{2}$ : $$CR_{1}(t)(i,j) = CR(t)(i,j) . \frac{cap(t)(i).cap(t)(j) }{\sum_{k=1}^{F}cap(t)(k)^{2}}~~ (19)$$
\item Indicator B3B: the spectral radius of the matrix $CR_{2}(t)$. Its coefficients are those of $CR(t)$ which have been weighted at each date $t$ by the volume of stocks exchanged  ($volu(t)$) expressed in dollars, in the following way for a dataset containing $F$ assets. $\forall (i,j) \in [1,F]^{2}$ : $$CR_{1}(t)(i,j) = CR(t)(i,j) . \frac{volu(t)(i).volu(t)(j) }{\sum_{k=1}^{F}volu(t)(k)^{2}}~~ (20)$$
\item Indicator B3C: Since indicator B3B will prove useful but will also usually produce a noisy signal, B3C is computed at each date $t$ as a moving average of B3B. We chose to average on 150 days, which is also the length $T$ of the rolling window. $$B3C(t) = \frac{\sum_{k=1}^{T}B3B(t-k)}{T} ~~ (21)$$
\end{itemize}
\end{itemize}

To conclude this section, we have therefore built nine financial crisis indicators in all. There are three of the first kind (A1, A2, A3) and six of the second kind (B1, B3, B3, B3A, B3B, B3C). Now we are going to use those indicators on the seven datasets that we possess in order to detect the periods of crisis.

\section{Empirical Results, Global Studies}
In this section we look at the global profiles produced by the nine financial crisis indicators that we have built for the seven datasets that we have chosen, and study in a qualitative manner what happens around the crises that we presented in Table 1. A more quantitative approach will be taken in the next section. Starting with the indicators of the A-series, we obtain the following results, the crisis events in Table 1 have been added as vertical purple lines and A1 is in blue, A2 in green and A3 in red. As a general remark, we can say that although useful global structures do clearly appear, all the profiles also appear to be quite noisy. We cannot miss the collapse of Lehman Brothers (Crisis 15) for example and it is a fact that most other crises are accompanied by very noticeable patterns in the value of the Hellinger distance, but there are also many false positives that blur the message of the indicators of the A-series. A1 and A2 always produce profiles that are similar, which is explained by the close resemblance between the reference distribution $\Theta1$ and $\Theta2$ (as seen in Figure 1 and in appendix), while the inverted indicator A3 produces radically different profiles. Many times we see A1 and A2 climb just before a truly major crisis, while A3 plummets. Indeed, the distribution $\Theta3$, which was derived from the covariance matrix obtained from a random matrix made of sums of very heavy tailed and correlated Student (t=3) coefficients is a good reference for a market that is in the process of dislocation, with the preeminence of very large eigenvalues in the covariance matrix that are indicative of dynamical instability. With that point of view, the correct way to read the profiles is: if A1 and A2 go up, then we are moving away in the sense of Hellinger from a distribution that is characteristic of a calm market, danger might be around the corner. If on top of that A3 is going down, then we are moving closer in the sense of Hellinger to a distribution of the eigenvalues of the covariance matrix that is characteristic of a market in distress. If those two effects are happening at the same time, then this pattern in the behavior of the indicators tends to indicate that the probability of a truly major market event is getting dangerously high.

\begin{itemize}
\item For Dataset 1, the pure international equity dataset, we get Figure 4 below. We observe elevated levels of A1 and A2 in the aftermath of Black Monday and during the build-up toward the S\&L crisis and Japanese Asset Price Bubble of 1990. Then there is a relative period of calm in the early 1990'. Monetary crises like Black Wednesday are not going to be visible using a dataset that does not contain any foreign exchange (FX) data because, despite causing a lot of pain, especially in the U.K, its long lasting influence on the global financial system remained limited. Then there is a sharp increase of A1 and A2 accompanied by a sudden characteristic drop of A3 just before and during the terrible blow of the 1997 Asian Crisis (Crisis 7). The 2000 NASDAQ crash is not very visible on those profiles, even though the NASDAQ is part of Dataset 1. Maybe this is due to the fact that it remained primarily an "American crisis" that is not going to be very apparent in a dataset that focuses primarily on contagion between international markets. The bullish market period of the early 2000's is characterized by mostly flat profiles of A1, A2 and A3 indicating a globally stable market structure. Before the Lehman Brothers collapse of 2008, we see again that pattern of an increase in A1 and A2 accompanied by a drop of A3.
\begin{center}
\includegraphics[width=150mm]{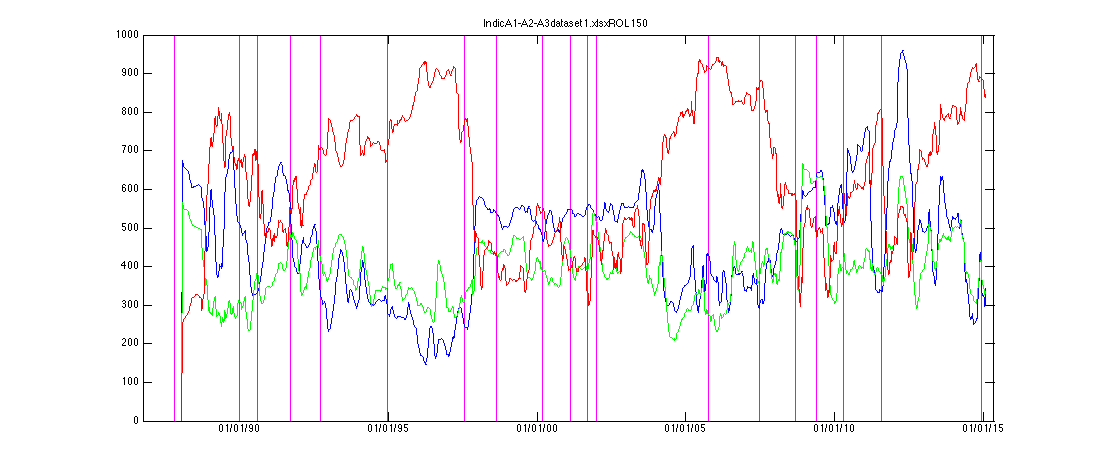} Figure 4: A1 blue, A2 green, A3 red
\end{center}

\item For Dataset 2, which is Dataset 1 augmented with commodities and safe, cash equivalent securities, we obtain the profiles below (Figure 5). Those profiles structurally resemble those obtained from Dataset 1 but they are much tamer, probably because of the presence of safe haven securities inside the dataset which provide a way for market agents to re-invest their money as they liquidate equity positions in the times leading up to and during a financial crisis. Since Dataset 2 includes commodities, we observe a very noticeable buildup in A1 and A2 leading up to the December 2014 Russian crisis (Crisis 19). It is again accompanied by an ominous drop in A3. The increased correlation between commodity and energy securities, which are represented in the equity indices by the major U.S oil companies, in the times leading up to a financial crisis, creates a strong build-up of market instability which provides valuable beforehand information that a crisis is becoming more likely.
\begin{center}
\includegraphics[width=150mm]{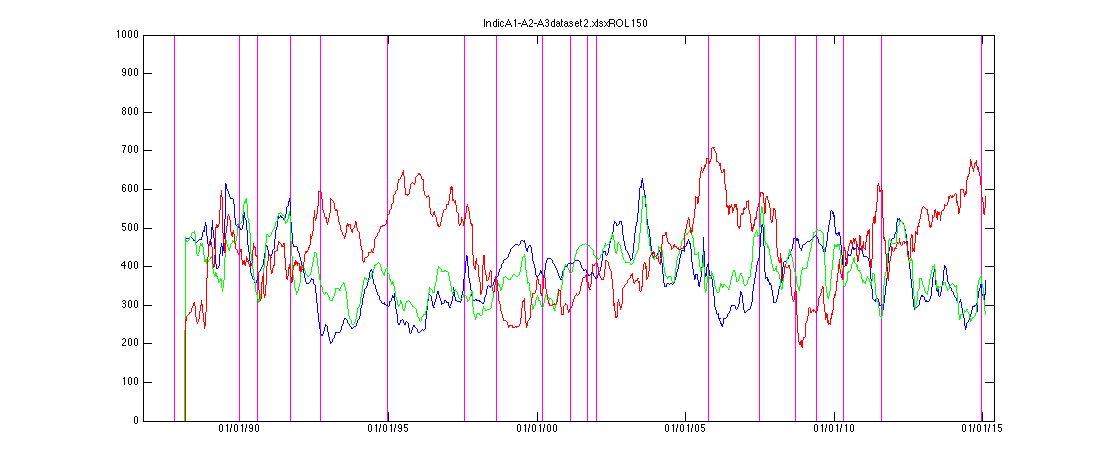} Figure 5: A1 blue, A2 green, A3 red
\end{center}

\item For Dataset 3, which is the "American" dataset, we get the following profiles (Figure 6).
Those profiles emphasize primarily, as anticipated, the events when the U.S market is overheating. The Savings and Loans crisis (Crisis 2) is anticipated a few months in advance by a sharp increase in A1 and A2. The indicators are mostly unresponsive during the Asian crisis of 1997 but the NASDAQ crash of March 2000 is this time accompanied by a spectacular spike of A1 and A2 accompanied by a depression in A3. This very good anticipation of the NASDAQ crash for this dataset which contains the sector components of the SP500 could be explained in part by the fact that the  information technology sector component of the SP500 is correlated at over 90\% with the NASDAQ. In the wake of the dot-com bubble, the same pattern reproduces itself around the 9-11 attacks. The same phenomenon happens again in the times leading up to the Sub-prime Crisis of August 2007 (although the drop in A3 is less noticeable) and on an even grander scale around the time of the Lehman Brothers collapse. The bankruptcy of General Motors on June 1st 2009 and the U.S sovereign credit degradation of August 5th 2011, when Standard \& Poor's reduced the country's rating from AAA (outstanding) to AA+ (excellent), are also anticipated by a spike in A1 and A2 while the behavior of A3 is less easy to interpret in those instances.
\begin{center}
\includegraphics[width=150mm]{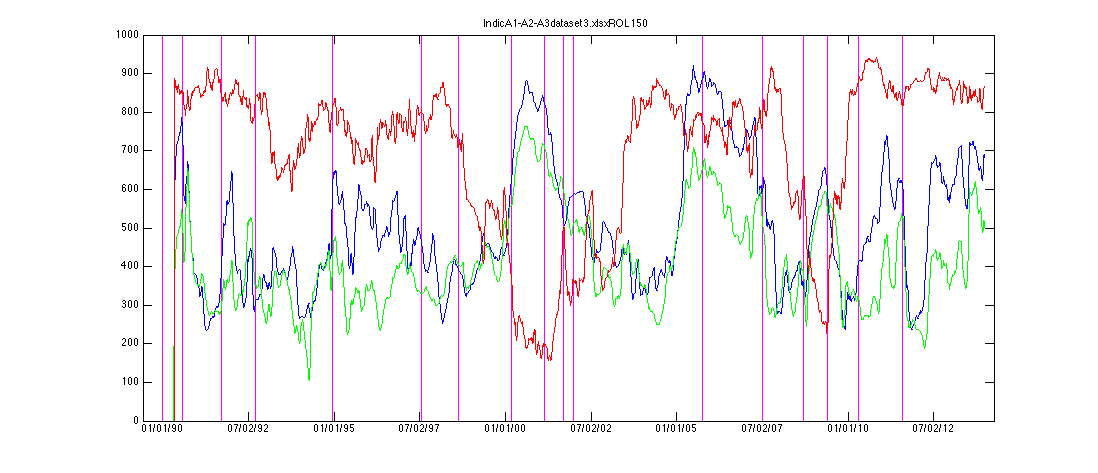}  Figure 6: A1 blue, A2 green, A3 red
\end{center}

\item For Dataset 4, which is the "European" dataset, we get the following profiles (Figure 7).
\begin{center}
\includegraphics[width=150mm]{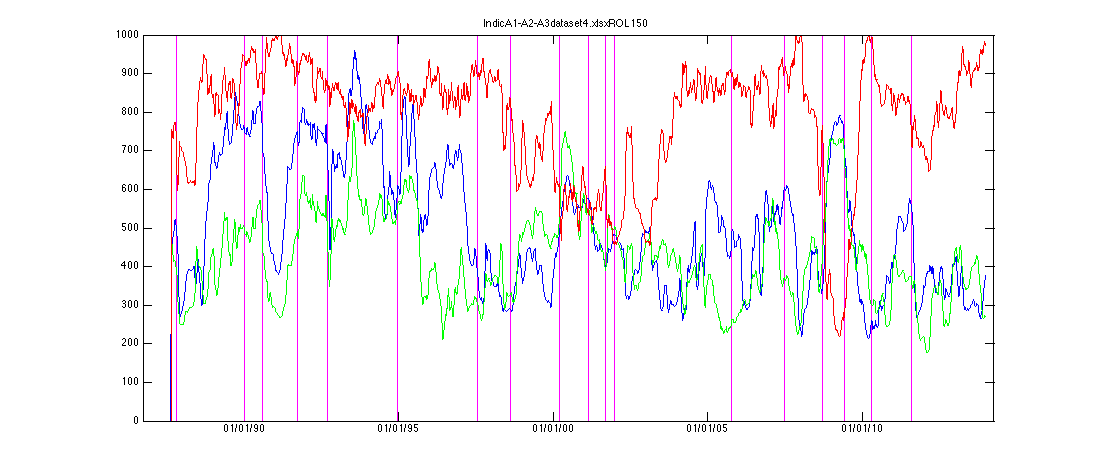}  Figure 7: A1 blue, A2 green, A3 red
\end{center}
Since the U.S financial market still leads the world, the global structure of the profiles of A1, A2 and A3 is somewhat similar to the structure of the profiles we had obtained for Dataset 1 and Dataset 3. There are however some very interesting specificities, which are characteristic of the European nature of Dataset 4. Events like the S\&L crisis, the NASDAQ bubble burst and even the Sub-prime Crisis of 2007 that preceded the Lehman Brothers collapse of 2008 are much less visible while the European sovereign debt crisis of April 23rd 2010 (Crisis 17) is spectacularly well anticipated with a huge spike in A1 and A2 accompanied by the ominous drop in A3. The U.S sovereign credit degradation of 2011 is also very well anticipated. That could be explained by the fact that the sovereign credit degradation of several leading European countries (France was degraded as well from AAA to AA1 by Moody's on November 19th 2012) was also being discussed by the media and anticipated by the financial markets.

\item For Dataset 5, the \textit{flight to quality} dataset which exploits the increasing correlation between equity and bonds in the times leading up to a financial crisis, we obtain the A1, A2 and A3 profiles below (Figure 8). In a way they seem even tamer and noisier than those of Dataset 2 which included, besides the commodities, some safe haven securities adding elements of \textit{flight to quality} to its design as well. In Dataset 5, the profiles of the indicators of A-series present few remarkable features, besides the obvious ones that all the others possess, like the spike in A1, A2 and the drop in A3 around the failure of Lehman Brothers in 2008. Unfortunately it seems that the inclusion of high quality sovereign and corporate bonds into the equity mix produced a dataset that is a little too resistant to most financial crises and is therefore, for the indicators of A-series at least, of limited interest as a way to anticipate crisis events.
\begin{center}
\includegraphics[width=150mm]{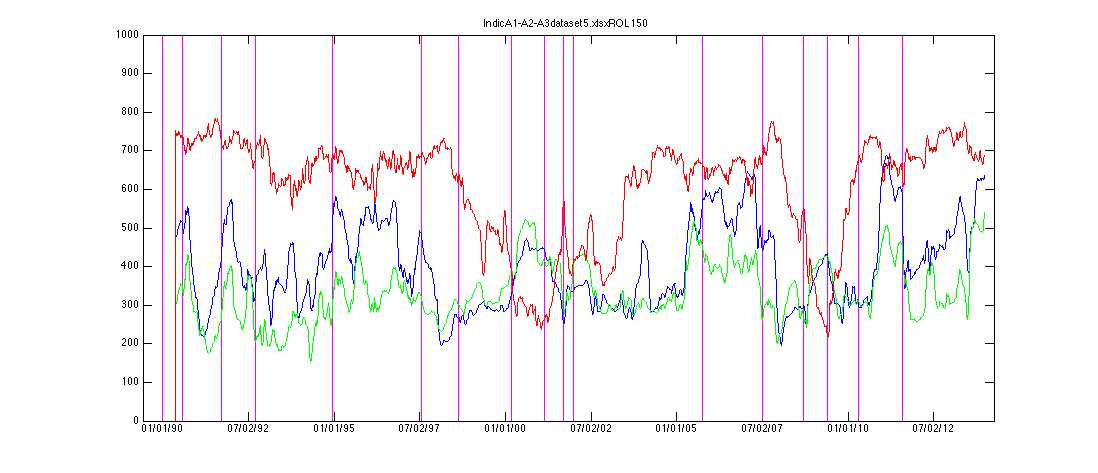} Figure 8: A1 blue, A2 green, A3 red
\end{center}

\item Dataset 6, the largest dataset containing 226 individual components of the SP500 index, produces the following high quality profiles when used with the indicators of the A-series (Figure 9). Black Monday is outside of the span of Dataset 6 and the S\&L crisis is surprisingly not visible (maybe it is because the affected companies dropped out of the index which, like all composite stock indices, shows some survivorship bias). The buildup to the 2000 NASDAQ crash is spectacular and characterized by the usual climb of A1 and A2 and fall of A3 as the correlations and volatility simmer inside the market and the spectrum of the covariance matrix is shifting to the right, toward the larger eigenvalues, away from $\Theta1$ and $\Theta2$ and toward $\Theta3$. This pattern started well in advance of the actual crisis event, which we chose to place on March 10th 2000 when the NASDAQ started its sharp and sustained fall. The indicators did provide a valuable early warning in that instance. Then we observe the period of bullish market in the early 2000's and again a slow buildup of correlations and volatility inside the market characterized by an increase in A1 and A2 (but not a drop in A3 that seems to only accompany truly catastrophic events) culminating during the Sub-prime Crisis of August 2007 which sets into motion the pattern of A1 and A2 spiking while A3 plummets leading up to the collapse of Lehman Brothers. Then there is some recovery in the market before the 2011 U.S sovereign credit rating degradation, which is anticipated by an increase in A1 and A2. Finally the Russian financial crisis of December 2014, which hit very hard some of the largest firms inside the SP500 (energy companies), because of the fall in the price of crude and gas, is anticipated by an increase in A1 and A2 but we don't observe much on A3 at the same time.
\begin{center}
\includegraphics[width=150mm]{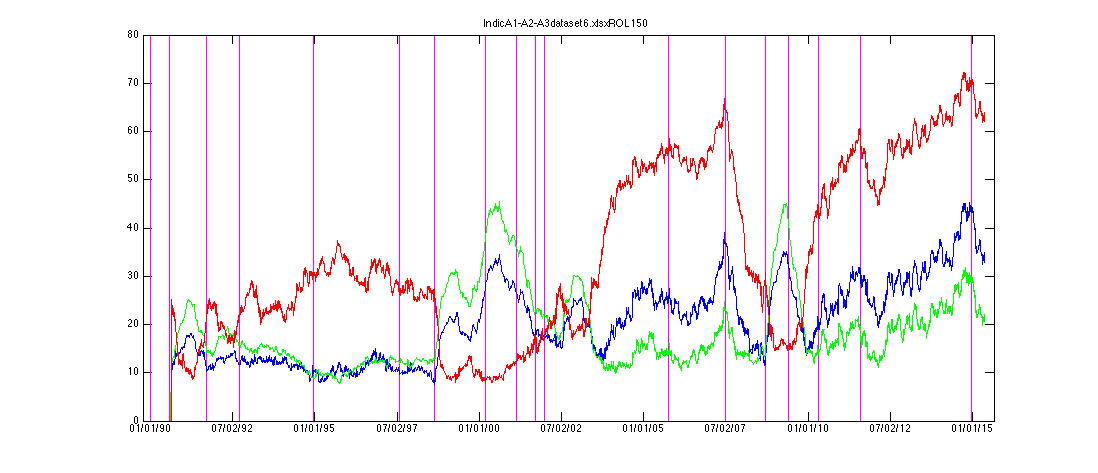} Figure 9: A1 blue, A2 green, A3 red
\end{center}
\item The A-series profiles for Dataset 7 are represented below (Figure 10). This dataset  contains international indices from emerging economies converted from the local currency into U.S dollars. The usual features that dominated all the other A-series profiles for all the other datasets look a bit diluted in the case of Dataset 7: the collapse of Lehman Brothers is one unremarkable bump among dozens of others and the NASDAQ crash is not visible, for example. The Delphi bankruptcy of late 2005 ("Red October") did trigger an economic crisis in many emerging countries due to the closure or expected closure of many of the overseas factories of the American automotive parts giant and this event is indeed anticipated by a rise in A1 and A2, but it is difficult to differentiate it from the many false positives. The Argentine sovereign default of late 2001 is surprisingly not visible although many South American indices (and the MERVAL itself) are included in Dataset 7. The Russian crisis of 2014 is however much more visible and better anticipated now than with the previous datasets and we observe a large spike in indicators A1 and A2 accompanied by a noticeable drop in indicator A3.
\begin{center}
\includegraphics[width=150mm]{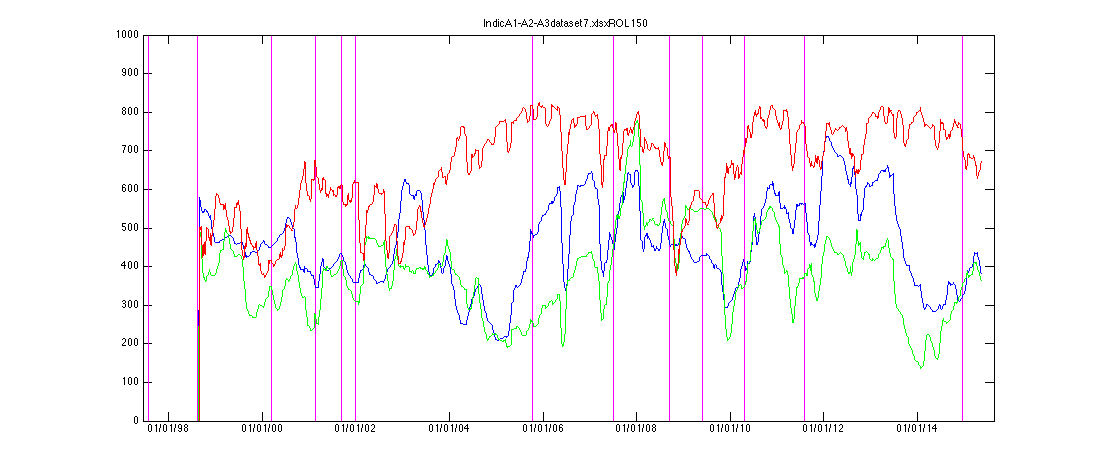} Figure 10: A1 blue, A2 green, A3 red
\end{center}
\end{itemize}

We now switch our attention to the indicators of the B-series. On all the profiles below, the crises of Table 1 are again represented as vertical purple lines. The spectral radius of the covariance matrix (mixed volatility and correlation signal), Indicator B1, is in green. The trace of the covariance matrix (volatility signal), Indicator B2, is in red. The spectral radius of the correlation matrix (correlation signal), Indicator B3, is in blue and is not represented on the same scale as the others for better readability (we multiplied it by 20). For Dataset 6, the correlation signal for the assets weighted by the market capitalization and volume traded, as defined in the previous section, is also going to be studied. We first remark that all the profiles of the B-series feature many false positives, as those of the A-series and the spikes are located, with specificities depending on the dataset used, in the vicinity of the crisis events of Table 1 for B1 and B2. The structure of the B3 profiles is usually a little  harder to interpret but it still holds valuable information.\\

For most datasets and most crises, B1 and B2 produce very similar profiles. When the profiles of B1 and B2 get closer to one another (ie. for the covariance matrix, the trace becomes close to the spectral radius) it means that correlations are increasing inside the financial market because one eigenvector's direction (the direction of the spectral radius) is becoming dominant over all the other ones. We do not however generally observe on the graphs that B3 is increasing when B1 and B2 are getting closer and that is due to the fact that B3 is only taking correlations into account while B1 is a mixed signal of volatility and correlation. The correlation component of B1 is at each date an average correlation weighted by the volatility of the assets constituting the dataset. In order to compare the relative position of B1 and B2 to the behavior of B3, we could have weighted the coefficients of the correlation matrix by the volatility of the assets. However doing this would have defeated the goal to study the correlations alone. In other words, when the relative position of B1 and B2 is not compatible with the behavior of B3, then it means that the assets are becoming correlated or uncorrelated depending on their volatility. For example when B1 and B2 are getting closer (correlations are increasing) but B3 is not increasing in a clear manner, it means that the high volatility stocks are becoming uncorrelated while the low volatility stocks are becoming correlated.\\

\begin{itemize}
\item For Dataset 1, we obtain the following results (Figure 11). This international equity dataset produces profiles of B1 and B2 which are very similar, with spikes in the vicinity of the major international crises like the NASDAQ crash and the failure of Lehman Brothers. The relaxation in both volatility and correlations after a crisis event is also much more noticeable than with the profiles based on the Hellinger distance, especially after the NASDAQ crash and the onset of the bullish market period. The B1 and B2 signals are almost always co-monotonic, it is what we expected and it is coherent with the work of Sandoval Junior and De Paula Franca (2012) who proved, using data which was very different from the data that we used, that high volatility in financial markets usually accompany a high level of correlations.
\begin{center}
\includegraphics[width=125mm]{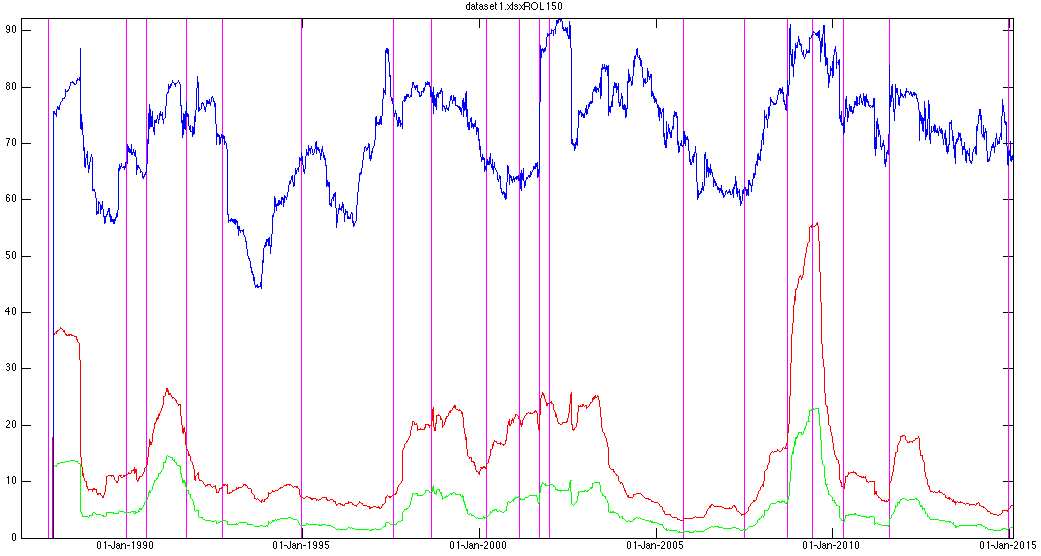}
\includegraphics[width=60mm]{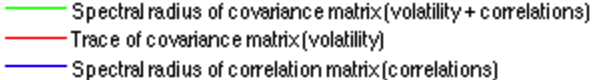} Figure 11: B1 green, B2 red, B3 blue (20x)
\end{center}

\item For Dataset 2, we obtain the following results (Figure 12). The profiles are very similar to those of Dataset 1, with less prominent features because of the inclusion of safe haven securities in the dataset. The elevation in both volatility and correlations is a little more noticeable around the Russian sovereign crisis of 1998, which may result from the inclusion of energy related commodities in this dataset. It also  highlights the typically increased correlations between energy (like oil companies stocks) and commodity securities in the times leading up to a financial crisis.
\begin{center}
\includegraphics[width=125mm]{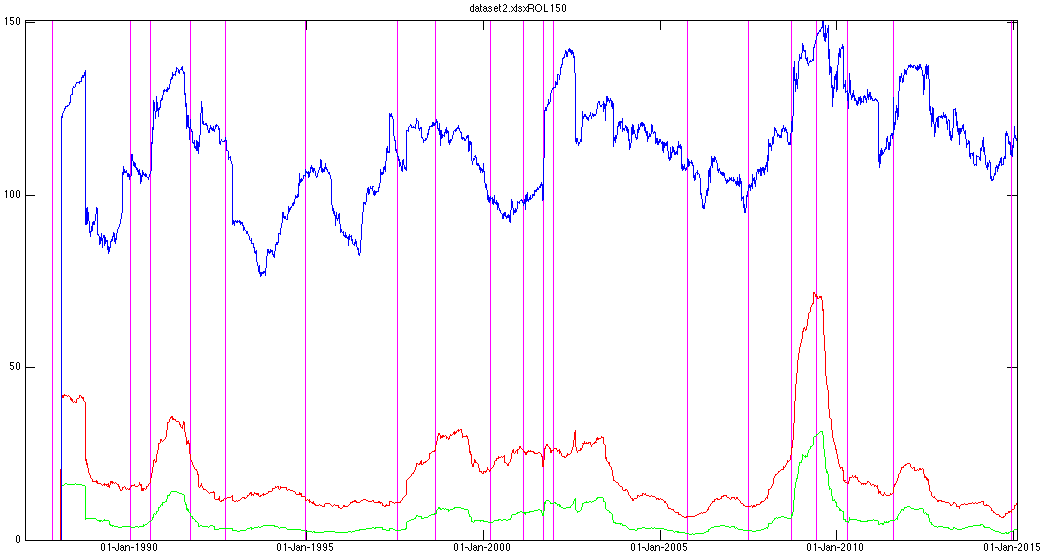}
\includegraphics[width=60mm]{supercartouche.png} Figure 12: B1 green, B2 red, B3 blue (20x)
\end{center}

\item For Dataset 3, we obtain the following results (Figure 13). The profiles of B1 and B2 highlight the American nature of Dataset 3. There are especially visible features for the NASDAQ crisis, the 2008 global financial crisis and the U.S sovereign credit degradation. The profile of B3 is again more difficult to interpret. In particular there is an apparent massive drop in correlations following the NASDAQ crisis that is not accompanied by a similar drop in volatility.
\begin{center}
\includegraphics[width=105mm]{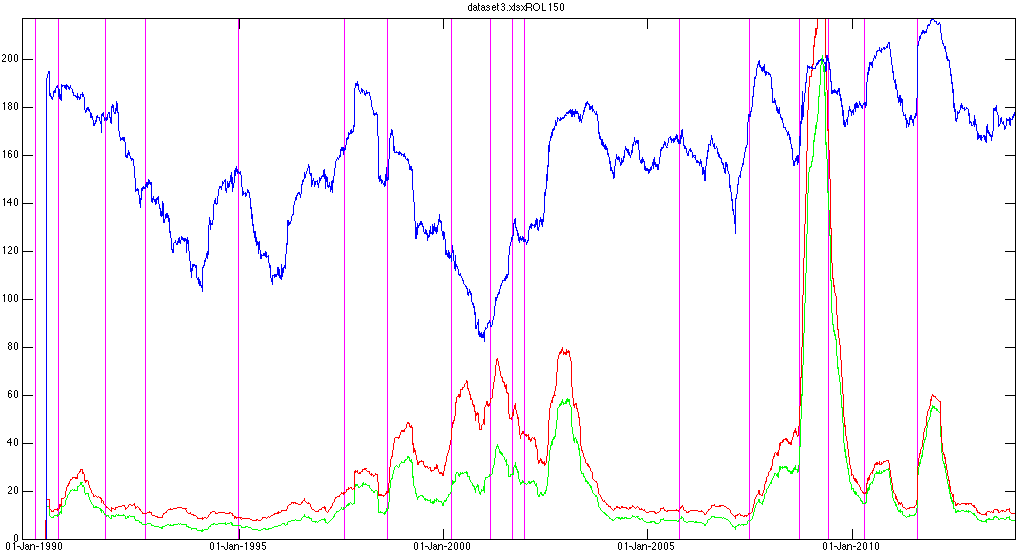}
\includegraphics[width=60mm]{supercartouche.png} Figure 13: B1 green, B2 red, B3 blue (20x)
\end{center}

\item For Dataset 4, we obtain the following results (Figure 14). It is the European counterpart of Dataset 3 and produces better detection of crises that are mostly or originally European in nature like the Eurozone sovereign debt crisis.
\begin{center}
\includegraphics[width=105mm]{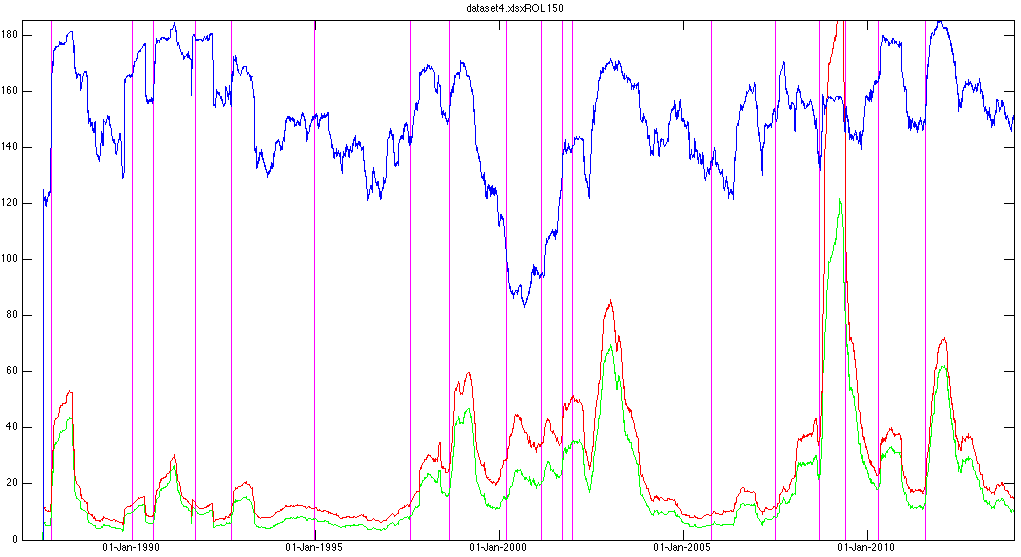}
\includegraphics[width=60mm]{supercartouche.png} Figure 14: B1 green, B2 red, B3 blue (20x)
\end{center}

\item For Dataset 5, we obtain the following results (Figure 15). The inclusion of high quality bonds to model the \textit{flight to quality} effect in the times leading up to a financial crisis has the effect, like for the A-series indicators, to produce very tame profiles of B1 and B2. Only the truly momentous events like the NASDAQ crash and the 2008 crisis are visible but all the other events are difficult to see, even the Asian crisis of 1997 which is surprising. The volatility signal B3 is still very noisy but some structure is starting to emerge with big spikes in the vicinity of known crises and large drops afterwards when the market is entering a post-crisis relaxation phase.
\begin{center}
\includegraphics[width=110mm]{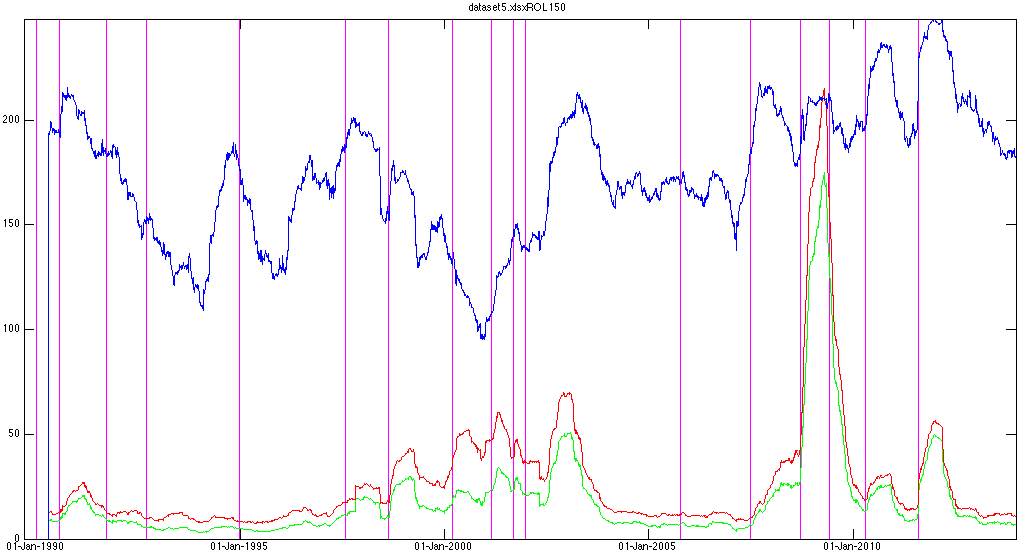}
\includegraphics[width=60mm]{supercartouche.png} Figure 15: B1 green, B2 red, B3 blue (20x)
\end{center}

\item For Dataset 6, we obtain the following results (Figure 16,17,18,19).
\begin{center}
\includegraphics[width=110mm]{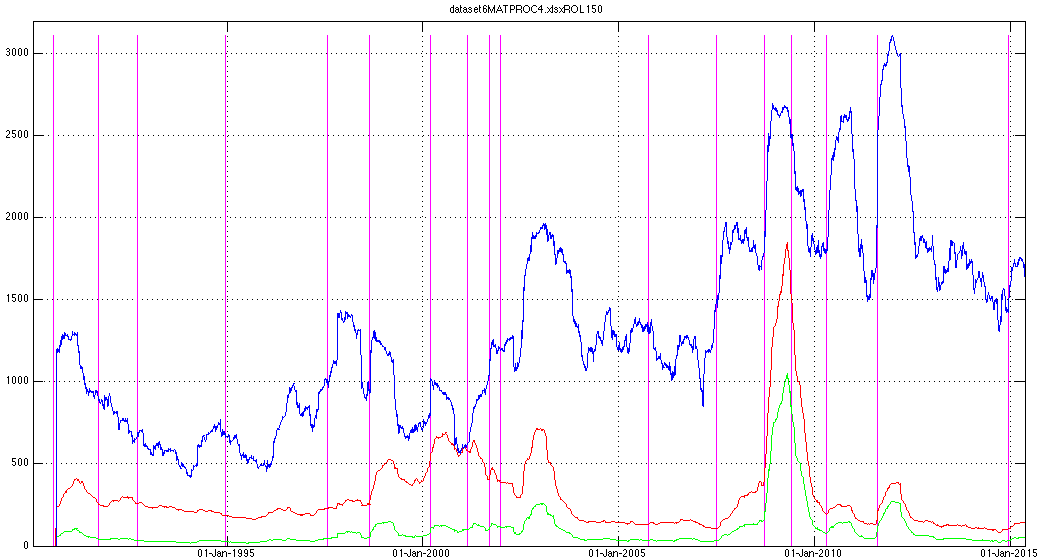}
\includegraphics[width=60mm]{supercartouche.png} Figure 16: B1 green, B2 red, B3 blue (20x)
\end{center}
The profiles for B1 and B2 are globally similar to those that we obtained for Dataset 1 and Dataset 3. However, the buildup of leverage during the sub-prime crisis is much more clearly visible and this time the correlation signal B3 in blue is much more interesting and contains a lot of usable information about the detection and anticipation of many of the financial crises of Table 1. Indeed, this increased precision of the results that we obtain is not surprising because Dataset 6 is constituted of a large number of individual stocks instead of indices and there is no averaging effect on the correlations.

\begin{center}
\includegraphics[width=110mm]{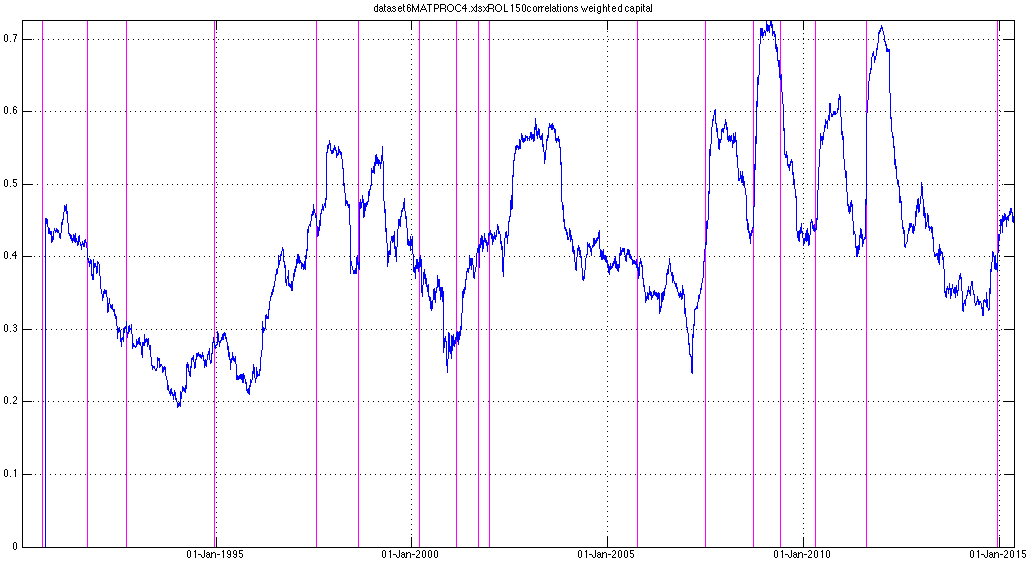}
Figure 17\\
 B3A : correlation signal B3 weighted by the capital of the companies corresponding to the stocks. We see new patterns emerge and a possible increase of the power of prediction for this indicator due to the weighting.
\end{center}

\begin{center}
\includegraphics[width=110mm]{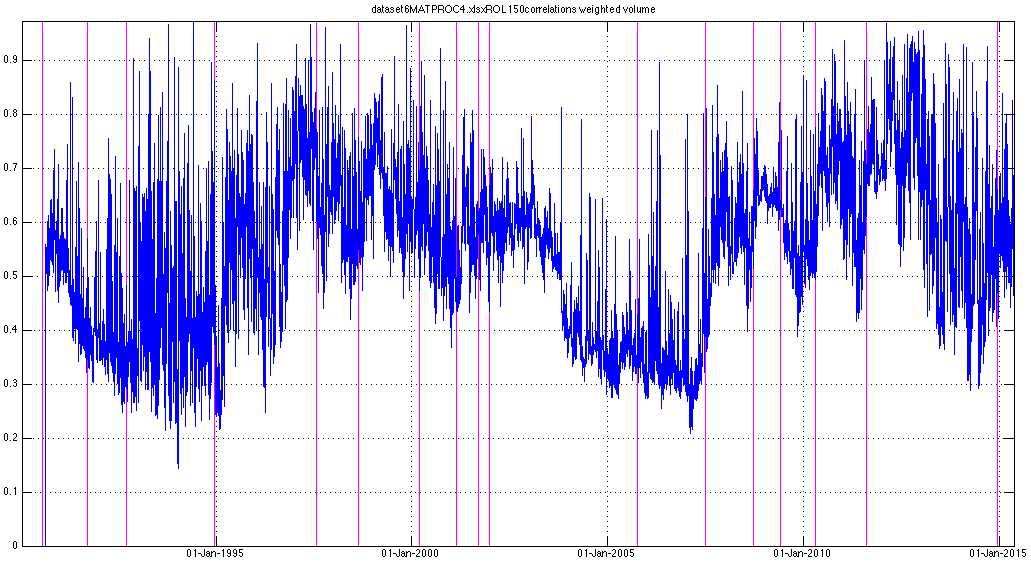}
Figure 18\\
 B3B : correlation signal B3 weighted by the volume traded. New patterns emerge and the power of this indicator to preempt rather than merely confirm the crises, as it was likely the case for most of the financial crisis indicators we have studied until now, appears to have increased.

\includegraphics[width=125mm]{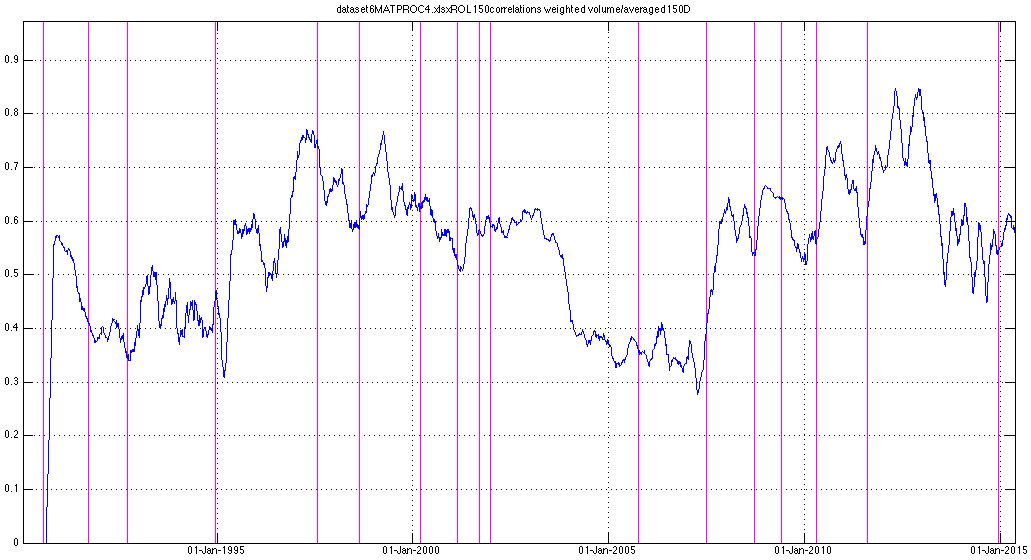}
Figure 19\\
 B3C : averaged version of B3B for better readability and noise reduction.
\end{center}

\item For Dataset 7, we obtain the following results (Figure 20).
\begin{center}
\includegraphics[width=125mm]{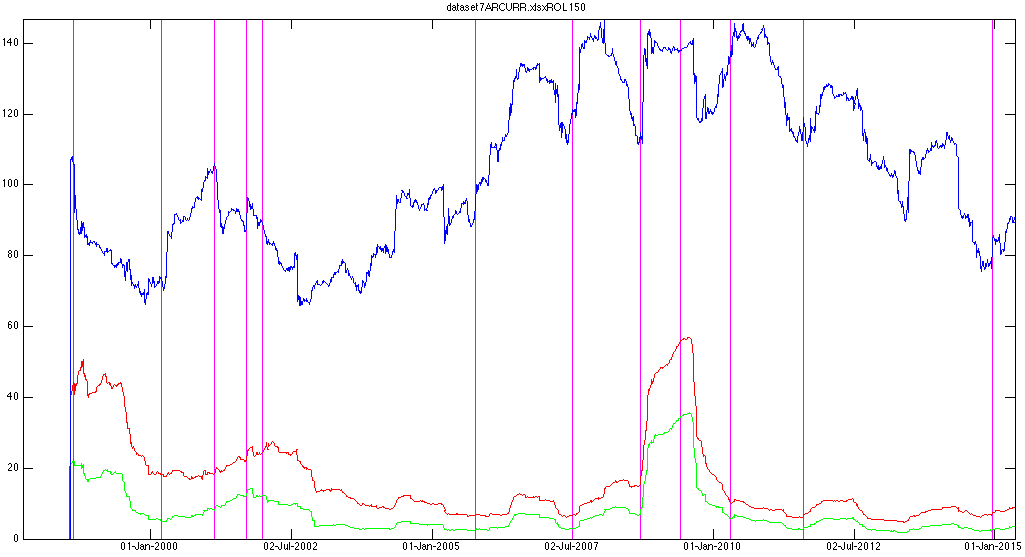}
\includegraphics[width=60mm]{supercartouche.png} Figure 20: B1 green, B2 red, B3 blue (20x)
\end{center}
\end{itemize}
The profiles do not seem to possess much useful signal and few truly noticeable features appear besides the obvious ones for B1 and B2 as well as for B3. The Asian crisis of 1997 (located at the very edge of the dataset) seems to have been much more visible when using Dataset 7 than when using the other datasets. This is without any doubt due to the fact that Dataset 7 contains information about emerging markets. Unfortunately in this study, we can only see the aftermath of the Asian crisis.

\section{Predictive Power}

In this section, we study in details the predictive power of indicators B3B and B3C, which are those that are based on the correlations weighted by daily traded volume of the assets. We decided to concentrate this study on those two indicators for simplicity. It is however clear that the same work could have been done using any of the nine indicators. We use Dataset 6, which contains the components of the SP500 index. The choice of this particular dataset is justified by the fact that it is the largest and most detailed of all the datasets we possess and therefore it is the one for which we expect to obtain the best quantitative results. We rely on the Maximum Draw Down (MDD) at horizon $H=100$ days, which is commonly used by practitioners in the financial industry. The reference asset price $A$ for which this quantity is computed at each date is the SP500 index.

$$MDD_{H}(t) = max_{t \leq x \leq y \leq t+H }  \left( 1-\frac{A(y)}{A(x)} \right) ~~ (22)$$ 

\subsection{Historical Approach}

We work with the scatter plots $[B3B(t), MDD_{H}(t)]$  and $[B3C(t), MDD_{H}(t)]$ for all dates $t$ covering the span of Dataset 6. We notice that the structure of each of those two scatter plots is dominated by a double threshold. We are going to exploit that fact in order to build a trigger for the indicators. At a given time $t$, when the value of B3B (resp. B3C) is between those twin thresholds, we say that we are in the \textit{danger zone}, which is where the probability of a crisis happening within 100 days is the highest. This makes also a lot of sense from a theoretical point of view. As a matter of fact, when the weighted correlations are low, then the probability of a crisis happening (i.e. experiencing a very high MDD over the next 100 days) is low, but when it is extremely high that means that we are already right in the middle of a crisis and the expected MDD at 100 days is low as well because the market would be likely out of the crisis and already in full recovery.\\

A crisis from Table 1 happening at time $t_{0}$ is (ex-post) considered to be predicted by B3B (resp. B3C) if at least 60\% of the points $[B3B(t), MDD_{H}(t)]$  (resp. $[B3C(t), MDD_{H}(t)]$)  such that  $t \in [t_{0}-100, t_{0}]$ are in the danger zone of the indicator. The false positive ratio is defined as the number of points inside the danger zone that belong to one of the crisis of Table 1 over the total number of points inside the danger zone.\\

To define the danger zone, we separate Dataset 6 into two periods: one in-sample calibration period between January 17th 1990 and December 31st 1999 and one out-of-sample forecast period where the power of prediction of B3B and B3C is going to be put to the test. For both indicators, the scatter plot restricted to the calibration period and the scatter plot covering the whole sample are represented below in Figure 21 and Figure 22. Both for B3B and B3C, the two scatter plots have roughly the same global structure. That fact is quite reassuring with regards to the validity of the approach that we have chosen. Indeed it means that whether we are considering the in-sample calibration period of the whole span of Dataset 6, the behavior of the indicators is the same and the danger zone is stable. This stationarity of the danger zone is crucial if we intend to make useful predictions in the out-of-sample forecast period.
\begin{center}
\includegraphics[width=51mm]{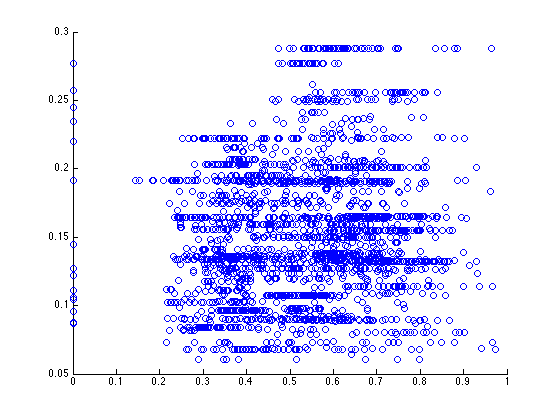} \includegraphics[width=51mm]{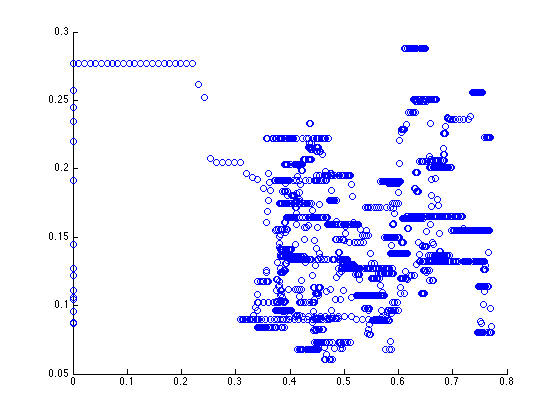} 
\end{center}
Figure 21: Calibration period / left: B3B, right: B3C

\begin{center}
\includegraphics[width=51mm]{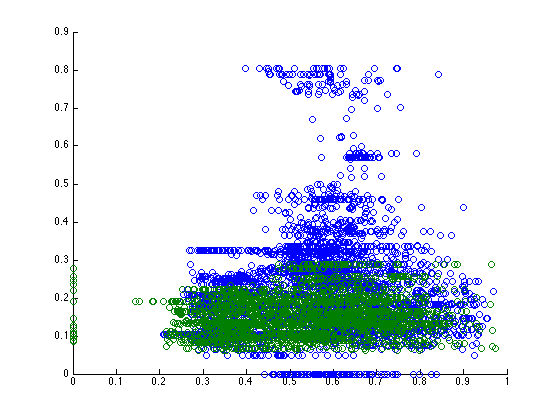} \includegraphics[width=51mm]{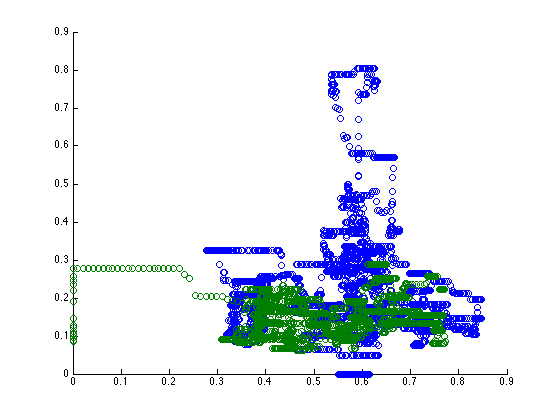} 
\end{center}
Figure 22: Calibration period: green, Whole sample: blue ; left: B3B, right: B3C\\

Using Figure 21, we set empirically the optimal values for the low and high thresholds that define the danger zones of the indicators. We obtained the intervals  [0.41,0.8] for B3B and [0.50,0.70] for B3C. Using those values, we are now able to count for each crisis of Table 1 inside the out-of-sample forecast period the number of dates in a 100 days window preceding each crisis for which the value of B3B (resp. B3C) was inside the danger zone as well as the global proportion of false positives.\\

We obtain the following results (Table 2) in which the crises inside the calibration period have been put in italic (Crisis 1 and Crisis 2 are located before the start of Dataset 6).

\begin{center}
\begin{tabular}{ | l | l | l | }
\hline
	&B3B & B3C  \  \\ \hline
	Crisis 1 & NA & NA \\ \hline
	Crisis 2 & NA & NA \\ \hline
	\textit{Crisis 3} & \textit{0} & \textit{0} \\ \hline
	\textit{Crisis 4} & \textit{51} & \textit{18} \\ \hline
	\textit{Crisis 5} & \textit{22} & \textit{0} \\ \hline
	\textit{Crisis 6} & \textit{41} & \textit{0} \\ \hline
	\textit{Crisis 7} & \textit{74} & \textit{0} \\ \hline
	\textit{Crisis 8} & \textit{98} & \textit{100} \\ \hline
	Crisis 9 & 93 & 100 \\ \hline
	Crisis 10 & 94 & 100 \\ \hline
	Crisis 11 & 98 & 100 \\ \hline
	Crisis 12 & 100 & 100 \\ \hline
	Crisis 13 & 13 & 0 \\ \hline
	Crisis 14 & 15 & 0 \\ \hline
	Crisis 15 & 98 & 100 \\ \hline
	Crisis 16 & 99 & 100 \\ \hline
	Crisis 17 & 97 & 100 \\ \hline
	Crisis 18 & 96 & 100 \\ \hline
	Crisis 19 & 90 & 78 \\ \hline
	False Positive (\%) & 73.38 & 71.30 \\ \hline
\end{tabular}
\end{center}
\begin{center}
Table 2: Historical Crisis Prediction
\end{center}

We observe that using the indicators B3B and B3C properly calibrated would have enabled us to predict and anticipate almost all the crises of Table 1 inside the out-of-sample forecast period. Indeed there are very few false negatives. Only the Bankruptcy of Delphi (Crisis 13) and the Sub-prime Crisis (Crisis 14) are not properly forecasted.  Those two events were however difficult to predict. Indeed, the effects of the Bankruptcy of Delphi were not instantaneous on the U.S economy in general and on the SP500 in particular and the Sub-prime Crisis was a months long process starting in August 2007 and on which we had to pin a date. The proportion of false positives, while still relatively high, is unlikely to represent an insurmountable obstacle for a practitioner as the information at a given date $t$ that there is around a 30\% chance of a crisis happening (and a 70\% chance of nothing happening) in the next 100 days is already an extremely valuable piece of information and a much more precise one than what we had expected initially. Indeed, in the introduction we where talking about a 10\% chance of a crisis happening being already a very valuable piece of information from an investor's point of view.

\subsection{Algorithmic Trading Approach}

In this section we do not consider the historical financial crisis events of Table 1 anymore. Instead, we focus on the MDD at each date, computed ex-post using the market data for the 100 days horizon. A financial crisis is  defined here as a market event where the MDD at 100 days exceeds a given threshold. We consider MDD thresholds from 10\% for a mild crisis to 25 \% for a serious downturn. This approach frees us of the sometimes arbitrary choice of a date for the financial crises of Table 1 and the signal given by the various indicators is now going to be made usable as the decision-making tool underlying an algorithmic trading strategy. In this study to showcase the predictive power of the indicators that we have built, we choose to consider Indicator B3B, but the same work could have be undertaken using any of the nine indicators of either the A-series or B-series. The complete construction of an optimal trading strategy based on those indicators will be the topic of an upcoming paper and in this section we merely intend to demonstrate that the red flags produced by B3B generate few false negatives, especially for higher MDD thresholds, and an acceptable proportion of false positives. The binary signal produced by one of the indicators (i.e. "red flag" \& "no red flag") can afterwards interact with a set of predetermined rules to give at each date a "\textit{buy}", "\textit{sell}" or "\textit{stay}" recommendation.\\

We separate Dataset 6 into two periods again: one in-sample calibration period and one out-of-sample period where the forecasting power of the financial  crisis  indicators that we have built is going to be tested. The period spanning from January 17th 1990 to December 29th 2000 is chosen this time for the calibration of the indicators. It is slightly larger than in the previous subsection and encompasses the 2000 NASDAQ crisis in order to boost the predictive power of the indicator by including a large market event in its calibration period (especially since the 1987 crash lies outside Dataset 6). We empirically chose the optimal interval [0.47,0.75] for the danger zone using Figure 23 below.\\

\begin{center}
\includegraphics[width=165mm]{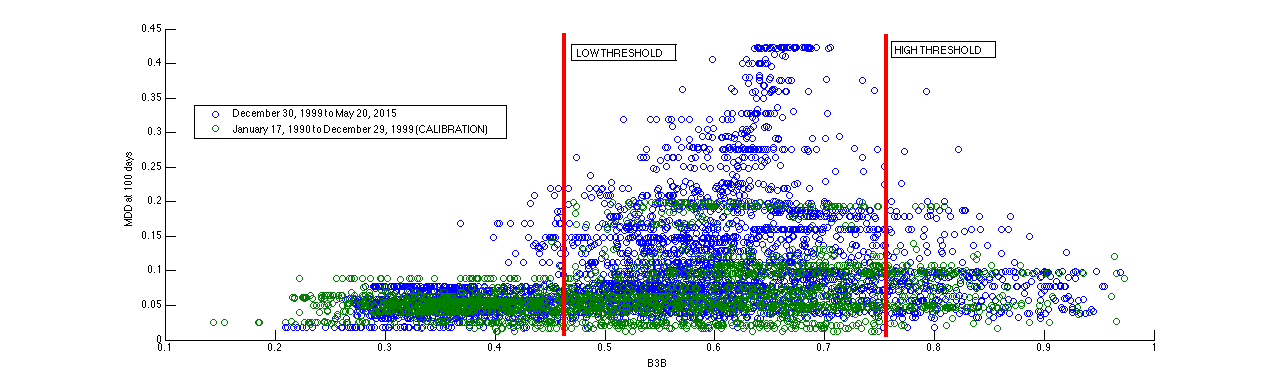}
Figure 23 (Optimal danger zone for B3B / Green: calibration period, Blue: forecast period)
\end{center}

We kept the rule that states that a red flag is given at a date $t_{0}$ if at least 60\% of the points $[B3B(t), MDD_{H}(t)]$ such that  $t \in [t_{0}-100, t_{0}]$ are in the danger zone of the indicator and we obtain the following results (Table 3) for various thresholds of MDD at the horizon 100 days. 

\begin{center}
\begin{tabular}{ | l | l | l | l | l | }
\hline
	MDD Threshold & 10\% & 15 \% &  20 \%& 25 \% \\ \hline
	Crises & 1281 & 794 & 400 & 268 \\ \hline
	Predicted crises & 1201 & 751 & 400 & 268 \\ \hline
	False positives & 1083 & 1533 & 1884 & 2016 \\ \hline
	False negatives & 80 & 43 & 0 & 0 \\ \hline
\end{tabular}
\end{center}
\begin{center}
(Table 3: Crisis Prediction)
\end{center}

The forecast period is constituted of 3770 trading days. For a MDD threshold of 10\% there are 1281 crisis events in the forecast period and Indicator B3B predicts 1201 of them while missing 80 and giving 1083 false positives. When the MDD threshold is put at 25\%, there are only 268 crisis events in the forecast period and Indicator B3B predicts all of them while giving 2016 false positives. The fact that Indicator B3B does not miss any crisis event characterized by the larger MDD threshold ($\geq 15 \%$) is reassuring from a financial stability point of view: an investor using a trading strategy based on the indicator that we have built would not have been caught off guard by a serious market downturn. Of course there are a lot of false positives, especially for the larger crises and there is room for improvement but, as we said in the introduction, the information that there is only a 10\% chance of a serious market event to happen within the next 100 days has tremendous value for market agents. False negatives can spell disaster while false positives might merely reduce profit and in that sense Indicator B3B is already very useful for a traditional risk averse investor or a regulator. However, if profit maximization is the most important benchmark for a less risk averse investor, then we could say that Indicator B3B performs best for the medium intensity crises (15 \% to 20 \% MDD threshold)  because it is there that the proportion of false positives remains smaller while the risk of a false negative is still acceptable. \\
\\
The predictive power of the financial crisis indicators developed in this study can also be demonstrated by considering European put options used as a protection tool against market downturns. The use of listed options for portfolio structuring had indeed become the norm among portfolio managers and many techniques exist to use those derivatives in order to decrease risk or boost performance, as explained in Bookstaber (1985). Bookstaber and Clarke (1981) also explained how using different kinds of put and call options in a portfolio helps set the necessary balance between risk and maximization of return.\\

Our purpose in this study is not to detail an elaborate structured portfolio but rather to demonstrate the predictive power of our financial crisis indicators, therefore we will limit ourselves to a basic protective-put strategy where the purchase and sale of European put options are merely used as a risk and volatility decreasing techniques. The performance of the various test portfolios, while still evidently an important benchmark, is not necessarily optimized like it would have been by, for example, considering more elaborate strategies, like a covered call instead of a protective put.\\

We work again with Indicator B3B and Dataset 6 because it is experimentally the setup that works the best and produces the most useful results, but the same study could have been conducted with any of our financial crisis indicators of either the A-series or the B-series. The basic idea is to compare the performances of three demonstration portfolios, which are updated monthly: 

\begin{itemize}
\item A static \textit{buy and hold} portfolio ($BAH$) constituted only of shares of an Exchange Traded Fund (ETF) replicating the SP500 (SPX).
\item A dynamic passive protective-put portfolio ($PPP$) constituted of a mix of equal proportions of SPX shares and SPX put options purchased every month as a protection.
\item A dynamic active protective-put portfolio ($PPA$) constituted either of a mix of equal proportions of SPX shares and SPX put options, or of SPX shares only, depending on the risk signal generated by indicator B3B. In this portfolio, the put options are only purchased as a protection when our financial crisis indicator forecasts a higher risk of a financial crisis happening withing a given time horizon. In this study, the time horizon for the predictive power of the financial crisis indicators is 100 days, corresponding to the 100 days horizon of the MDD that we consider and which was used to experimentally obtain the danger zone of the indicator, as we have detailed previously.
\end{itemize}

The decisions dates are chosen every month as the third Friday of the month between January 2000 and May 2015. Indeed, the third Friday of every month is the day that put options with a maturity of one month or more traditionally reach maturity. For accurate option quotes, we use a database of real historical option prices for the SPX \footnote{ The date was purchased from the website \textit{historicaloptiondata.com}}.\\

The process of decision on whether or not to buy a protective put option in $PPA$ is the following: at a given monthly decision date $t_{0}$, we count the number $N_{t_{0}}$ of times when indicator B3B was inside its danger zone, as defined previously, in the 100 days preceding $t_{0}$.\\

The choice of an appropriate threshold $\mathscr{T}$ above which the value of $N_{t_{0}}$ triggers the purchase of the put option is crucial. If $\mathscr{T}$ is too high, we will not buy the protection at dates when it would have been prudent, but on the other hand if $\mathscr{T}$ is too low, we will buy many very expensive put options at dates when the risk of a crisis was relatively low according to our indicator, thus destroying the anticipated performance gain of  $PPA$ with respect to $PPP$ and $BAH$.\\

After careful considerations, we choose  $\mathscr{T}=80$. Such a value provides a good balance between the need to maximize the performance of $PPA$ and the need to keep its risk withing reasonable bounds.\\

At $t_{0}$, if $N_{t_{0}} \geq 80$ we interpret this signal as an indication of a heightened probability of a crisis happening in the near future, thus triggering the purchase of a put option for protection. On the other hand, if $N_{t_{0}} <80$, then we interpret this as a reassuring signal indicating that the probability of a crisis happening in the near future is relatively low and thus that the money needed to buy a put option for protection can be saved for increased performance of the portfolio.\\

The investors considered are \textit{price takers}. Unlike a market maker, they have to accept the prices that the market is offering them. Therefore, at a given monthly decision date, any put option is bought at its \textit{last ask price} and any put option is sold at its \textit{last bid price}, as provided by our database of real historical SPX option quotes. The price of the underlying (the SPX shares) is always taken as the last price of the day.
\\

More precisely, once a strike $S$ and a maturity $M$ for the protective put option has been chosen, and we will discuss about that below, the execution of our three test strategies will be as follows:\\

\begin{itemize}
\item The investor holding $BAH$ starts with a given number of SPX shares (normalized at 1 share in our computations) and holds on to it for the duration of the experience, which spans 185 months between January 2000 and May 2015. 
\item The investor who holds $PPP$ starts with a given number of SPX shares and purchases every month an equal number of put options (normalized at 1 in our study), while selling the put option already present in the portfolio. All the purchased options are kept only for one month and then they are sold again to help cover the cost of the next option purchase. Therefore, every month, the investor who holds $PPP$ buys a put option of strike $S$ and maturity $M$ and sell the option purchased the month before, which is also of strike $S$ and maturity $M$.\\

We do not include a cash allocation in our portfolios in order to boost performance and in order to be always entirely invested in the market. Besides, as explained in Harper (2003), the inclusion of cash in a portfolio is generally counterproductive and, regardless of performance consideration, a riskless security like U.S bonds is usually a better choice than plain cash anyway. Since there is no cash in our study, the SPX shares, which are very liquid, are used as cash-equivalent to purchase the options and the proceedings of the option sales is immediately converted into SPX shares as well. Disregarding the small technical detail of having to  consider fractional SPX shares in the computations, we rebalance $PPP$ at each decision date such that the proportions of SPX shares and put options remain equal (i.e there is always one unit of put option covering one unit of share). 

\item The investor who holds the dynamic active portfolio $PPA$ adopts a strategy with rolling put options to protect the SPX shares, similar in nature the strategy governing the $PPP$ portfolio, but with one fundamental difference: at a given decision date, the decision to buy, or not to buy, the protective put option is made according to the signal produced by indicator B3B. If at a given decision date the financial crisis indicator recommends not to buy the protective put option, then the investor sells the option purchased the month before, if it is present, does not buy any other option and reverts to a portfolio entirely made of SPX shares, like in the case of $BAH$.
\end{itemize}

After the choice of the threshold $\mathscr{T}$ that establishes the desired balance between performance and safety in the active strategy, the choice of the strike $S$ and the maturity $M$ of the rolling put options is the next important step that will determine whether the active strategy $PPA$ is a success or not.\\

The maturity $M$ has to be long enough so that when the put option is sold again after one month its value has not depreciated too much, but on the other hand, in order for our study to have meaning, we should not take the maturity so far away in the future that it goes beyond the horizon of prediction of the financial crisis indicator, which is fixed at 100 trading days and which represents, assuming 252 trading days per year, a little under five months. While taking this into consideration, we chose the maturity for the rolling put option equal to four months: $M = 4 \ months$\\

The strike $S$ is chosen at 100\% of the price of the underlying SPX share (put option \textit{at-the-money}). Our goal is to beat, both in terms of performance maximization and risk minimization, the static SPX portfolio $BAH$ with our active, financial crisis indicator controlled, portfolio $PPA$. Therefore, it is best not to settle for a loss mitigation approach by choosing a strike $S < 100$. The risk has also to be kept between reasonable bounds, but choosing to use more expensive at-the-money options did not in practice increase the volatility of the portfolios, while it did maximize overall performance.\\

For every decision date $t \in \llbracket 1,185 \rrbracket$, the riskless asset $TB$ is chosen as the one month U.S Treasury Bond (US0001M). Considering an asset $A$, which can represent either $BAH$, $PPP$ or $PPA$, we define the following benchmarks, Sharpe ratio and Calmar ratio, that we will use to compare the strategies to one another and demonstrate the predictive power of Indicator B3B. We also recall the Maximum Draw Down (MDD), that we have already defined in Equation (22). Here we consider MDD(A), which is the maximum draw down computed over the entire period of study and not anymore at a given time horizon.\\

\begin{itemize}

\item To compute the yearly Sharpe ratio, we proceed in the following manner. The Sharpe ratio measures the quotient of the excess performance with respect to a riskless asset over the volatility. To achieve the desired highest Sharpe possible, a strategy has to maximize return while at the same time minimize volatility, which represents risk.\\

\begin{itemize}
\item $\forall t \in \llbracket 1,185 \rrbracket$ , $ExcessReturn_{A}(t) =\frac{A(t)}{A(t-1)} - \frac{TB(t-1)}{12} -1$
\item $Perf(A) = (\frac{A(185)}{A(1)})^{\frac{360}{5550}} - 1$ (Annualized performance, assuming 30 days per month and 185 months in our experience)\\

\item $Vol(A)=stdev(ExcessReturn_{A}).\sqrt{12}$ (Annualized volatility)\\

\end{itemize}
$Sharpe(A) = \frac{Perf(A) - mean(TB)}{Vol} $ (23)\\

\item We define the yearly Calmar ratio as the quotient of the performance by the maximum draw down (MDD). \\

$Calmar(A) = \frac{Perf(A)}{MDD(A)}$ (24)
\end{itemize}

In Figure 24, we draw the profiles of $BAH$, $PPP$ and $PPA$ as well the quantity $TR=\frac{PPA}{BAH}$, which represents the extra performance of the active strategy governing portfolio $PPA$ with respect to the static portfolio $BAH$ (the tracking error). The Performance, the Sharpe and the profile of $TR$ are the main tools at our disposal to illustrate the benefit of using $PPA$ instead of all the alternatives presented in this study and thus demonstrate the power of prediction of our financial crisis indicator B3B, and by extension the power of prediction of our original approach to financial crisis indicators as a whole.\\

We immediately notice that $PPP$ is a complete failure. It has a negative Sharpe (anti performance with respect to the riskless asset) and while it does somewhat reduce volatility and MDD with respect to BAH, which is a good thing and which was expected given the very nature of the protective-put strategy, the complete collapse of its performance under the crippling cost of having to buy a new put option each and every month makes this strategy very unattractive. The cost of having to buy the protection every month is not to be underestimated. Indeed, as explained in Israelov and Nielsen (2015), the cost of buying the options in a protective-put setting is usually very high. It is also often much higher than it might seem, even during a calm market and low volatility period, unless the price and fundamental value of the underlying are properly taken into consideration.\\

The strategy PPA is a success, both in terms of maximization of the return and in terms of reduction of the risk. The performance of BAH was only 2.6\% and is boosted to almost 6\% in PPA, while the volatility goes down from 18\% to 13\% and the MDD goes down from almost 50\% to 33\%. These MDD values in particular show that the effects of the September 2008 financial crisis, with the failure of Lehman Brothers in particular, had been correctly anticipated by our financial crisis indicator, allowing the strategy PPA to anticipate the fall by buying protective put options in advance while also anticipating the post crisis recovery and stop buying the protection to save money when it was no longer necessary. The Sharpe ratio of PPA is 0.323 while it was only 0.047 for BAH, demonstrating a tangible gain of performance for our dynamic active protective-put portfolio PPA and therefore demonstrating the power of prediction of our financial crisis indicator.\\

While a Sharpe in the order of magnitude of 0.3 would still be considered modest from the point of view of a hedge fund manager, it must be pointed out that it is the result of a single and simple protective-put strategy, without any diversification. In a real-world setting, diversification of the assets in the portfolio and of the strategies as well as the choosing of more elaborate and realistic rules for the purchase of options, which could also include call options to finance the purchase of the put in a covered call framework, might very well produce much more impressive Sharpe ratios. Also the Sharpe ratio that we have obtained for PPA is computed over a very long period of 15 years that includes long phases of market stagnation as well as several major financial crises, which reduce the overall annualized performance. If we had truncated our study to make it start from 2007, for example, the Sharpe ratio of PPA would have been much larger. The Calmar ratio for its part goes from 5.2\% in BAH up to 18\% in PPA and demonstrates that using our financial crisis indicator to pilot a protective-put strategy permits to both increase performance and at the same time reduce the MDD, which is a very desirable outcome for an asset manager.\\

The study of the profile of the $TR$ coefficient is interesting as well. Like we have said it measures the extra performance of PPA with respect to BAH. Since the performance of $TR$ is 3.3\% it means that PPA is performing 3.3\% above the SP500 ETF, which is already a very good result. The structure of the $TR$ profile over time is remarkable as well. It features a sequence of increasing plateaux corresponding to the times of low market risk when Indicator B3B correctly recommends not buying the protective put option in the PPA strategy. Those plateaux, besides showing that our financial crisis indicator correctly anticipates most of the periods of low risk calm market, as well as anticipating the crises of course, shows that a correct selection of the threshold $\mathscr{T}$ at 80\% instead of 60\% enables us to limit the occurrences of false positives, without of course eliminating them and they still remain the main limitation in our framework. The Sharpe of $TR$, at around 0.278, while the Sharpe of BAH was only 0.047, demonstrates also the added value that the predictive power of our financial crisis indicator brings to portfolio PPA.\\

We now switch our attention to Figure 25 and Figure 26 which show the comparison between PPA and two kinds of random strategies. The basic idea between comparing PPA to random strategies is to show that the success of PPA compared to $BAH$ and $PPP$ is \textit{not} just due a stroke of luck and that the added value and predictive power of our framework of financial crisis indicator is real. In the first kind of random strategy, the choice to buy, or not, the protective put option is random at each one of the 185 monthly decision dates in our study. We just flip a coin at each date and decide accordingly whether to buy the protection or not. We call those strategies \textit{random at every date} (RED) strategies and they are shown in Figure 25. In the second kind of random strategies, that we call \textit{random with the same proportion} (RSP), one random strategy is equivalent to performing a random permutation on the signal that Indicator B3B provides for PPA ('1' for \textit{buy} and '0' for \textit{don't buy}). Those strategies are compared to PPA in Figure 26.\\

We observe that indeed, both the RED and RSP strategies perform on average much worse than PPA, which is very reassuring. They have on average higher volatility, much lower performance and a significantly larger MDD.  Visually, PPA performs better than most random strategies of both kinds because it is "above" most of them and it is located, most of the time, at the top of the area of the plane defined by the superposition of all the random paths. PPA is not in the middle, which would have suggested than on average the probability of doing better than PPA by adopting a random strategy would have been around 50\%, which would have seriously damaged the credibility of our approach. PPA is not at the bottom of the area defined by the random paths either, which would have been even worse and implied that our active strategies, on average, would have been beaten by random ones. In fact, while considering the 2000 simulated random strategies, either RED or RSP,  the strategy PPA beats them, in the sense that the final value of the PPA portfolio is above the final value of the portfolio governed by the random strategy, 1992 times for RED and 1989 times for RSP. That means that PPA is better in terms of global return than a random strategy around 99.5\% of the time. Portfolio PPA perform even better when compared to a RED strategy than to a RSP strategy since the RSP strategy, while random, do contain a little information about the signal provided by Indicator B3B in the form of the global proportion of purchases over the course of all the 185 decision dates. Computing the average Sharpe ratio of the random strategies confirms that the good performance of PPA is due to skill rather than luck. Indeed, while the Sharpe of PPA is 0.323, the average Sharpe ratio of the 2000 paths of RED and RSP is very close to zero(-0.0101 and +0.0034, respectively), which is still better than the Sharpe of PPP, for which the cost of buying the protection every month destroys the performance of the portfolio.

\begin{landscape}
\begin{figure}[p!]
\centering
\includegraphics[scale=0.59]{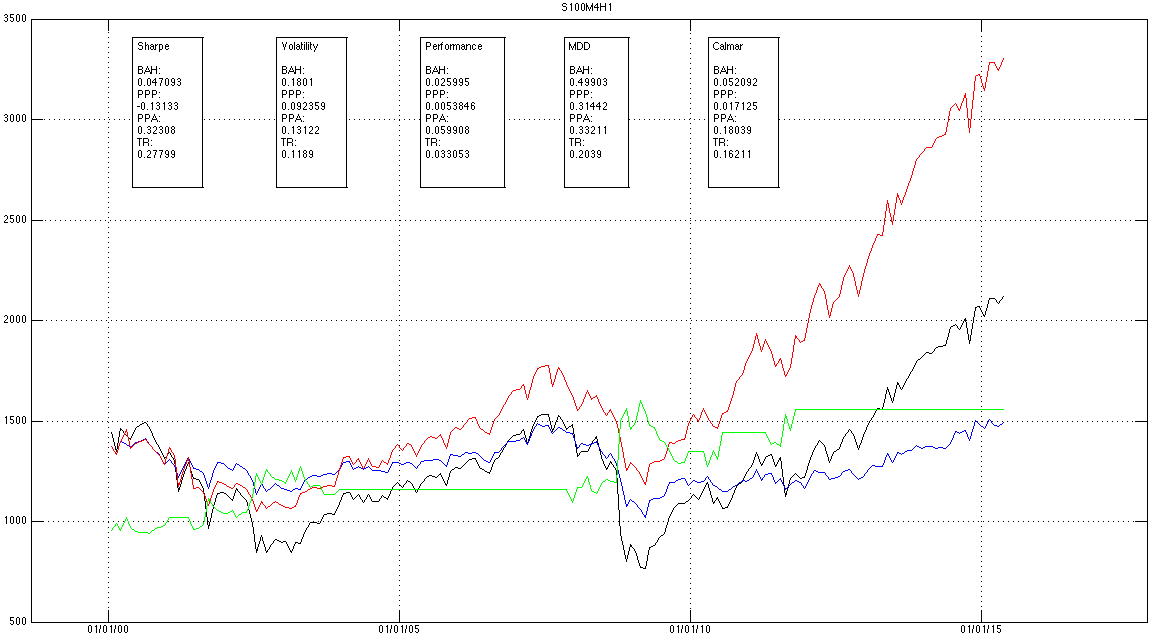}
\captionsetup{labelformat=empty}
\caption{Figure 24 (Value of the portfolios $BAH$ (black), $PPP$ (blue) and $PPA$ (red). The extra performance of $PPA$ with respect to $BAH$ (tracking error) is represented in green and is not on the same scale (x1000).)}
\end{figure}
\end{landscape}

\begin{landscape}
\begin{figure}[p!]
\centering
\includegraphics[scale=0.59]{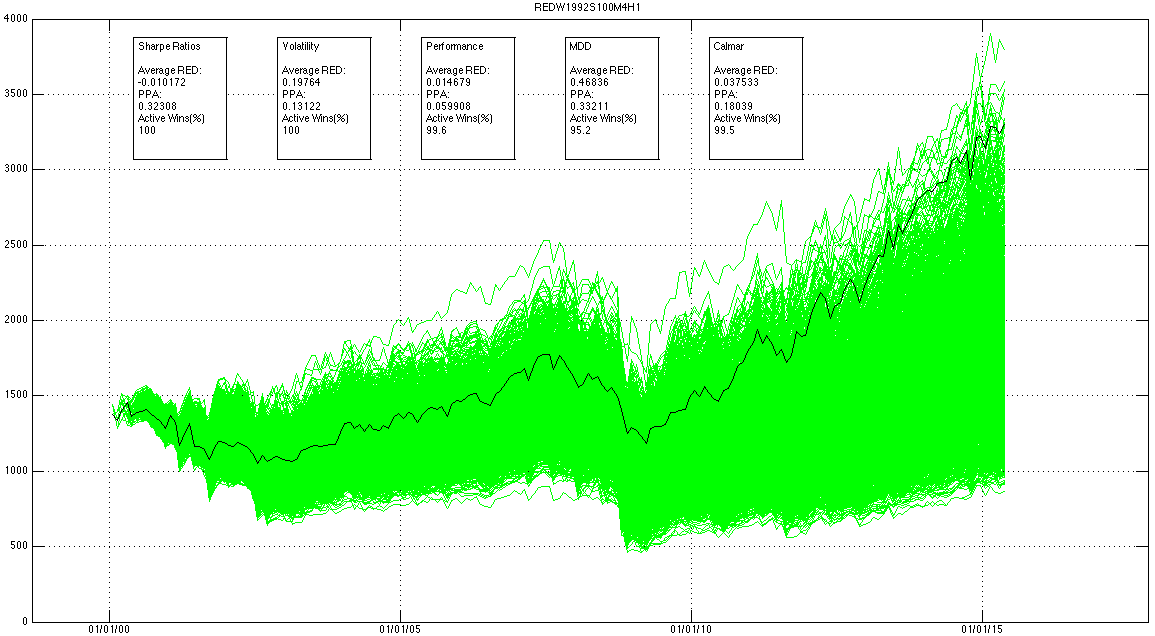}
\captionsetup{labelformat=empty}
\caption{Figure 25 (Comparison between $PPA$ (black) and 2000 random strategies (green) for which the choice of whether to buy, or not, the put option is random (coin flip) at every monthly decision date (random every date).)}
\end{figure}
\end{landscape}

\begin{landscape}
\begin{figure}[p!]
\centering
\includegraphics[scale=0.59]{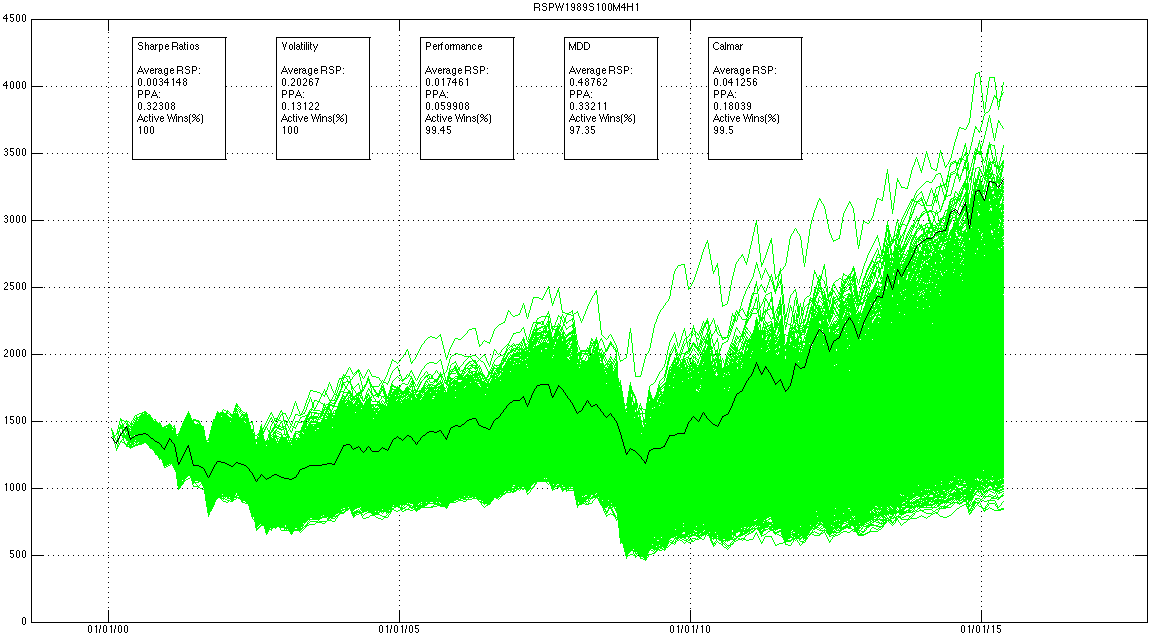}
\captionsetup{labelformat=empty}
\caption{Figure 26 (Comparison between $PPA$ (black) and 2000 random strategies (green) for which the choice of whether to buy, or not, the put option is random at every monthly decision date and the same global proportion of purchases as in $PPA$ over the 185 decision is preserved (random same proportion).)}
\end{figure}
\end{landscape}

\section{Conclusion}
As a general conclusion, we could start by saying that the nine financial crisis indicators that we have built are all generally able to detect most of the financial crises that we have studied. In both the historical approach, where we made use of chosen dates for the crisis events, and the algorithmic trading approach, where we used a more quantitative definition of the financial crises based on the MDD, the indicators were indeed capable of confirming the occurrence of market turmoil. Moreover, we demonstrated an out-of-sample predictive power for several of those indicators, while using a dataset constituted of selected components of the SP500 index. We also demonstrated the predictive power of the indicators in our framework by using a signal produced by Indicator B3B in order to build a successful portfolio constituted of SPX shares and European put options and governed by an active protective-put strategy.\\

We recall that we have built two sets of financial crisis indicators and that we then applied them on seven datasets. The financial crisis indicators that we have built are all based of the study of the spectrum of a covariance matrix, a correlation matrix or a weighted correlation matrix. They measure the volatility and correlations between a number of assets in order to evaluate whether the conditions are right for adverse random events, which are happening all the time, to trigger financial crises.\\

The first kind of indicators, that we called the A-series, comprises three indicators. Indicator A1 and A2 are at each date the Hellinger distance between the empirical distribution of the spectrum of the covariance matrix and two different calm market reference distributions. Indicator A3 for its part is at each date the Hellinger distance between the empirical distribution of the spectrum of the covariance matrix and a reference distribution characterizing a market in turmoil. We found that one of the most useful patterns for financial crisis detection and forecast in the profiles of the indicators of the A-series is characterized by a spike in A1 and A2 accompanied by a drop in A3. Indeed, when this pattern occurs, it means that the market is in the process of moving away from a calm state and toward more turbulence.\\

We called the indicators of the second type the B-series. Indicator B1 is the spectral radius of the covariance matrix and bases its forecasts on a mixed signal of volatility and correlation. Indicator B2 is the trace of the covariance matrix and relies on volatility only to make its predictions. Indicator B3 is the spectral radius of the correlation matrix and relies on correlation only to make its predictions. We also have built three additional versions of B3. B3A is the spectral radius of a correlation matrix where the assets have been weighted with regards to the market capitalization of the firms they represent. B3B is the spectral radius of a correlation matrix where the assets have been weighted with regards to their daily traded volume and B3C is an averaged version of B3B. We found that the indicators of the B-series, especially those that rely, in part or in whole, on correlation performed better while using the components of an index rather than a basket of indices. That probably has to do with the averaging effect that is a very strong influence in the computation of an index. Used on Dataset 6, which contains the components of the SP500, Indicator B3B is the one which gave us the best and the most reproducible results.\\

In the last part of the study, we demonstrated that Indicators B3B and B3C do possess, after proper calibration, a real out-of-sample power of prediction in estimating the probability of a financial crisis happening at a given time horizon in the future. While the approach that we adopted gave many false positives (a red flag is returned and no crisis happens in the market) as well, the low number of  false negatives reinforced our conviction about the viability and usefulness of the financial crisis indicators that we have built. For indicator B3B, we also developed a quantitative approach relying on defining a financial crisis in terms of the crossing of an MDD threshold. Indicator B3B was also successfully used in order to build a signal governing a protective-put strategy in a portfolio constituted of a mix of options and ETF shares. \\

We are confident that our framework and the financial crisis indicators that we have built are able to bring new insight on the topic of financial crisis detection and prediction. In the future, one of the possible ways of applying these methods would be to use them in order to build more elaborate systematic trading strategies, which make use of all our nine indicators as well as new ones. Those strategies based on the aggregated signals coming from many different financial crisis indicators in our framework will be the topic of an upcoming paper.

\begin{small}
 
\end{small}

\section*{Appendix}

\subsubsection*{Reference Distributions for all Datasets}    	
\begin{figure}[H]
\centering
\subfloat[Dataset 2]{
  \includegraphics[width=70mm]{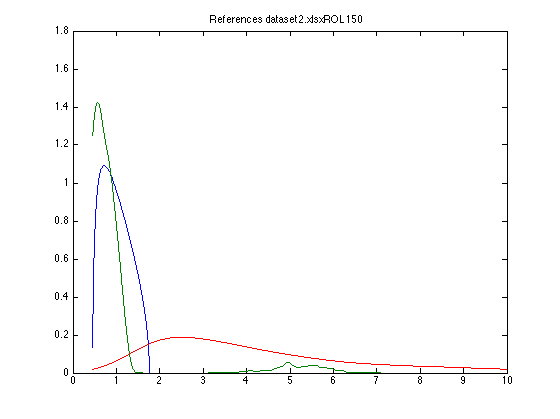}
}
\subfloat[Dataset 3]{
  \includegraphics[width=70mm]{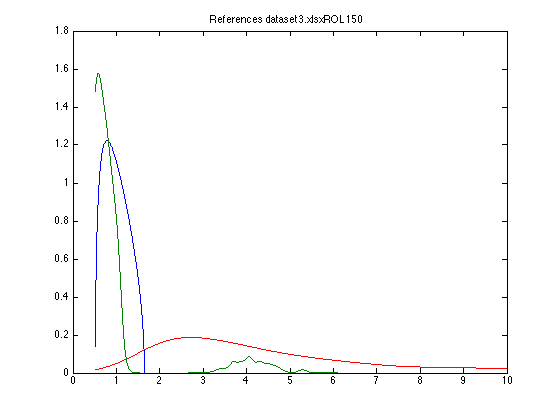}
}
\hspace{0mm}
\subfloat[Dataset 4]{
  \includegraphics[width=70mm]{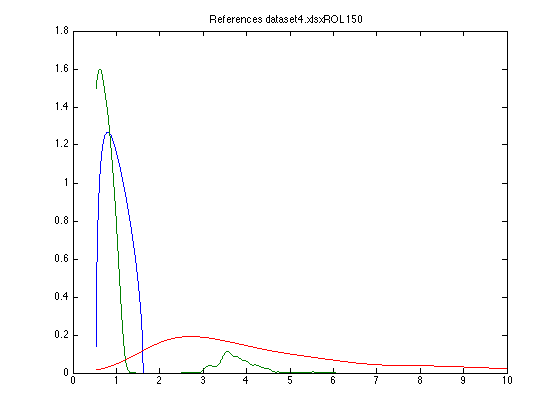}
}
\subfloat[Dataset 5]{
  \includegraphics[width=70mm]{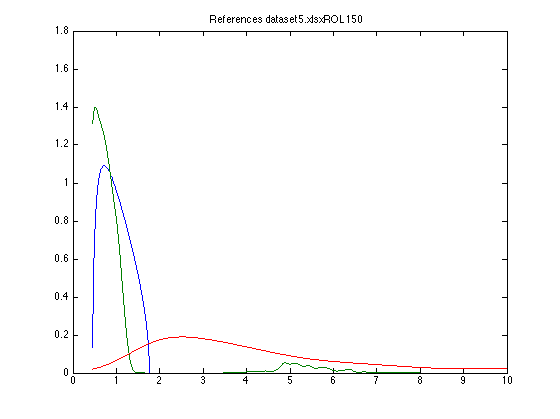}
}
\hspace{0mm}
\subfloat[Dataset 6]{   
  \includegraphics[width=70mm]{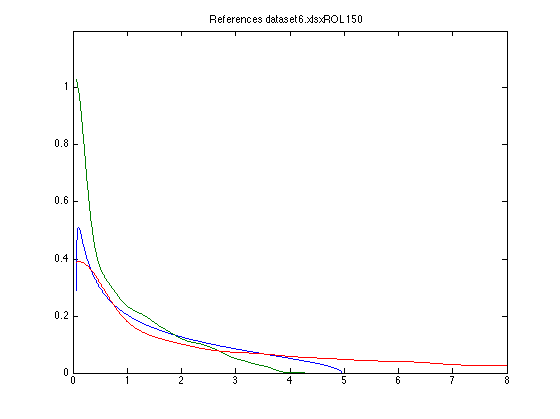}
}
\subfloat[Dataset 7]{
  \includegraphics[width=70mm]{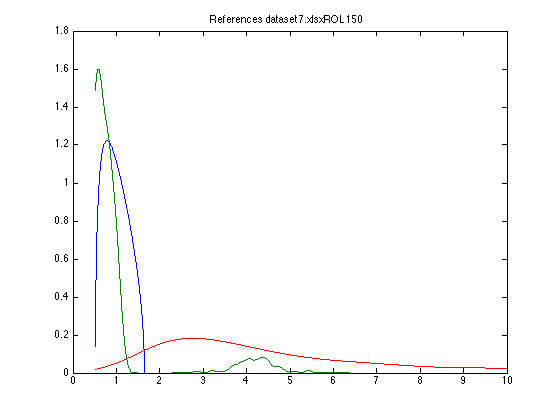}
}
\captionsetup{labelformat=empty}
\caption{Reference distributions computed for Dataset 2 to Dataset 7, with a rolling window of 150 days (Blue: Marchenko Pastur ($\Theta$1), Green: $\Theta$2, Red: $\Theta$3).}
\end{figure}

\subsubsection*{Composition of Dataset 6 (226 assets)}
AA,  
    AAPL,  
    ABT,  
    ADM,  
    ADSK,  
    AEP, 
    AFL,  
    AIG,  
    ALTR,  
    AMAT,  
    AMGN, 
    AON,  
    APA,  
    APC,  
    APD,  
    ARG,  
    AVY,  
    AXP,  
    BA,  
    BAC,  
    BAX,  
    BBY,  
    BCR,  
    BDX,  
    BEN,  
    BHI,  
    BK,  
    BLL,  
    BMY,  
    C,  
    CAG,  
    CAT,  
    CB,  
    CCE,  
    CELG,  
    CI,  
    CINF,  
    CL,  
    CLX,  
    CMA,  
    CMCSA,  
    CMI, 
    CMS,  
    CNP,  
    COP,  
    CPB,  
    CSC, 
    CSX,  
    CTAS,  
    CTL,  
    CVS,  
    CVX,  
    D,  
    DD,  
    DE,  
    DHR,  
    DIS,  
    DOV,  
    DOW,  
    DTE,  
    DUK,  
    EA,  
    ECL,  
    ED,  
    EFX,  
    EIX,  
    EMC,  
    EMR,  
    EOG,  
    EQT,  
    ES,  
    ETN,  
    ETR,  
    EXC,  
    EXPD,  
    F,  
    FAST,  
    FDO,  
    FDX,  
    FISV,  
    FITB,  
    FMC,  
    GAS,  
    GCI,  
    GD,  
    GE,  
    GIS,  
    GLW,  
    GPC,  
    GPS,  
    GWW,  
    HAL,  
    HAR,  
    HBAN,  
    HCP,  
    HD,  
    HES,  
    HOG,  
    HON,  
    HOT,  
    HP,  
    HPQ,  
    HRB,  
    HRL,  
    HRS,  
    HSY,  
    HUM,  
    IBM,  
    IFF,  
    INTC,  
    IP,  
    IPG,  
    IR,  
    ITW,  
    JCI,  
    JEC,  
    JNJ,  
    JPM,  
    K,  
    KLAC,  
    KMB,  
    KO,  
    KR,  
    KSU,  
    L,  
    LB,  
    LEG,  
    LEN,  
    LLTC,  
    LLY,  
    LM,  
    LNC,  
    LOW,  
    LRCX,  
    LUK,  
    LUV,  
    MAS,  
    MCD,  
    MDT,  
    MHFI,  
    MMC,  
    MMM,  
    MO,  
    MRK,  
    MSFT,  
    MSI,  
    MUR,  
    NBL,  
    NEE,  
    NEM,  
    NI,  
    NOC,  
    NSC,  
    NTRS,  
    NUE,  
    NWL,  
    OKE,  
    OXY,  
    PAYX,  
    PBCT,  
    PBI,  
    PCAR,  
    PCG,  
    PCL,  
    PCP,  
    PEG,  
    PEP,  
    PFE,  
    PG,  
    PGR,  
    PH,  
    PHM,  
    PKI,  
    PNC,  
    PNW,  
    PPG,  
    PPL,  
    PVH,  
    R,  
    ROST,  
    RTN,  
    SCG,  
    SHW,  
    SIAL,  
    SLB,  
    SNA,  
    SO,  
    SPLS,  
    STI,  
    SWK,  
    SWN,  
    SYMC,  
    SYY,  
    T,  
    TE,  
    TEG,  
    TGT,  
    THC,  
    TIF,  
    TJX,  
    TMK,  
    TMO,  
    TROW,  
    TRV,  
    TSO,  
    TSS,  
    TXT,  
    TYC, 
    UNM,  
    UNP,  
    USB,  
    UTX,  
    VAR,  
    VFC,  
    VMC,  
    VZ,  
    WEC,  
    WFC,  
    WHR,  
    WMB,  
    WMT,  
    WY,  
    XEL,  
    XOM,  
    XRAY,  
    XRX

\subsubsection*{Funding}    
This work was achieved through the Laboratory of Excellence on Financial Regulation (Labex RéFi) supported by PRES heSam under the reference ANR­10­LABX­0095. It benefited from a French government support managed by the National Research Agency (ANR) within the project Investissements d'Avenir Paris Nouveaux Mondes (Investments for the Future Paris ­New Worlds) under the reference ANR­11­IDEX­0006­02.
\end{document}